\documentclass[pdflatex,sn-nature]{sn-jnl}

\usepackage{graphicx}
\usepackage{multirow}
\usepackage{amsmath}
\usepackage{amssymb}
\usepackage{amsthm}
\usepackage{mathrsfs}
\usepackage[title]{appendix}
\usepackage{color}
\usepackage{xcolor}
\usepackage{textcomp}
\usepackage{manyfoot}
\usepackage{booktabs}
\usepackage{algorithm}
\usepackage{algorithmicx}
\usepackage{algpseudocode}
\usepackage{listings}
\usepackage{makecell}

\usepackage{titlesec}
\usepackage{threeparttable} 
\newcommand{\sfmark}{\textcolor{blue}{$^{\mathcal{SF}}$}}   
\newcommand{\uncmark}{\textcolor{red}{$^{\mathcal{U}}$}}     
\newcommand{\bothmark}{\textcolor{blue}{$^{\mathcal{SF}}$}\textcolor{red}{$^{\mathcal{U}}$}} 

\begin{document}
	
\title{Uncertainty-Guided Dual-Domain Learning for Reliable Skin Lesion Segmentation}
	
\author[1,2,3]{\fnm{Duwei} \sur{Dai}}
\author[2,4]{\fnm{Caixia} \sur{Dong}}
\author[5]{\fnm{Guowei} \sur{Dai}}
\author[6]{\fnm{Qingsen} \sur{Yan}}
\author[1]{\fnm{Qin} \sur{Zhang}}
\author[4]{\fnm{Fan} \sur{Liu}}
\author[4]{\fnm{Pengyu} \sur{Ren}}
\author*[2]{\fnm{Guangyao} \sur{Kong}}\email{konggy@xjtu.edu.cn}
\author*[2,3]{\fnm{Wei} \sur{Zeng}}\email{wz@xjtu.edu.cn}

\affil[1]{\orgdiv{Department of Dermatology}, \orgname{the Second Affiliated Hospital of Xi'an Jiaotong University}, \orgaddress{\city{Xi'an}, \postcode{710004}, \state{Shaanxi}, \country{China}}}

\affil[2]{\orgdiv{National-Local Joint Engineering Research Center of Biodiagnosis \& Biotherapy}, \orgname{the Second Affiliated Hospital of Xi'an Jiaotong University}, \orgaddress{\city{Xi'an}, \postcode{710004}, \state{Shaanxi}, \country{China}}}

\affil[3]{\orgdiv{School of Mathematics and Statistics}, \orgname{Xi'an Jiaotong University}, \orgaddress{\city{Xi'an}, \postcode{710049}, \state{Shaanxi}, \country{China}}}

\affil[4]{\orgdiv{Institute of Medical Artificial Intelligence}, \orgname{the Second Affiliated Hospital of Xi'an Jiaotong University}, \orgaddress{\city{Xi'an}, \postcode{710004}, \state{Shaanxi}, \country{China}}}

\affil[5]{\orgdiv{College of Computer Science}, \orgname{Sichuan University}, \orgaddress{\city{Chengdu}, \postcode{610065}, \state{Sichuan}, \country{China}}}

\affil[6]{\orgdiv{School of Computer Science}, \orgname{Northwestern Polytechnical University}, \orgaddress{\city{Xi'an}, \postcode{710072}, \state{Shaanxi}, \country{China}}}

\abstract{
Accurate skin lesion segmentation is the cornerstone of dermoscopic Computer-Aided Diagnosis. However, the intrinsic visual ambiguity and morphological irregularity of lesions---often stemming from diffuse pathological transitions---frequently defeat pure spatial modeling, prompting a shift toward multi-domain architectures. Yet, existing multi-domain paradigms suffer from a systemic oversight: the inadequate active utilization of prediction uncertainty. By relegating uncertainty to a passive, post-hoc metric, these models are trapped in a deterministic framework, manifesting as the ``blind fusion'' of cross-domain artifacts, ``passive'' topological reasoning, and ``rigid optimization'' that overfits to subjective label noise. To dismantle these bottlenecks, we propose the Uncertainty-Guided Dual-Domain Network (UGDD-Net), an uncertainty-in-the-loop unified architecture. Powered by a novel two-pass ``Glance-and-Gaze'' forward mechanism, UGDD-Net transforms uncertainty into an active guiding signal that permeates the entire learning lifecycle. Specifically, an Uncertainty-Guided Bi-directional Feature Fusion (UGBFF) module leverages pixel-level uncertainty to dynamically modulate spatial-spectral interactions, enabling high-confidence representations to rectify unreliable features. Furthermore, an Uncertainty-Guided Graph Refinement (UGGR) module constructs a priority-guided topology-aware graph to perform active global reasoning, propagating reliable semantic consensus to refine uncertain nodes. Finally, an Uncertainty-Guided Margin-Adaptive Loss (UGML) enforces strict constraints on confident pixels while relaxing penalties on uncertain ones, achieving appropriately cautious predictions and superior statistical calibration. Extensive experiments on four public datasets (ISIC2017, ISIC2018, PH$^2$, and HAM10000) demonstrate that UGDD-Net establishes a new state-of-the-art, with its superiority particularly pronounced on our curated ``Hard Samples'' subset. Moreover, our uncertainty maps exhibit meaningful alignment with expert inter-observer variability on diagnostically ambiguous images, providing robust interpretability for human-machine collaborative diagnosis.
}
	
\keywords{Skin lesion segmentation, Reliable prediction, Dual-domain learning, Uncertainty-guided learning, Topological reasoning}

\maketitle
	
\section{Introduction}\label{sec1}
	
Skin diseases rank among the leading causes of global disease burden, posing a pervasive threat to public health \cite{langselius2025global}. For aggressive malignancies like melanoma, early detection is decisive in reducing mortality; the five-year survival rate exceeds 90\% with early diagnosis but plummets below 15\% in advanced stages \cite{hrvatin2026completed}. Currently, dermoscopic evaluation relies heavily on manual analysis. However, the global shortage of dermatologists and the inherent diagnostic complexity of skin lesions render manual evaluation highly subjective, inevitably leading to severe inter-observer variability and potentially fatal treatment delays. Consequently, Computer-Aided Diagnosis (CAD) systems---offering objective, reproducible, and high-throughput analysis---have emerged as indispensable clinical tools. Accurate skin lesion segmentation serves as the cornerstone of CAD, providing critical quantitative morphological metrics for downstream clinical strategies, such as surgical margin assessment, severity grading, and personalized medication formulation. 	While traditional image processing techniques \cite{celebi2019dermoscopy} laid the initial groundwork, the intrinsic visual complexities of skin lesions render their rigid, hand-crafted rules largely ineffective. Consequently, this representational bottleneck has catalyzed a paradigm shift toward robust, data-driven deep learning methodologies.

\begin{figure}[!ht]
	\centering
	\includegraphics[width=0.6\linewidth]{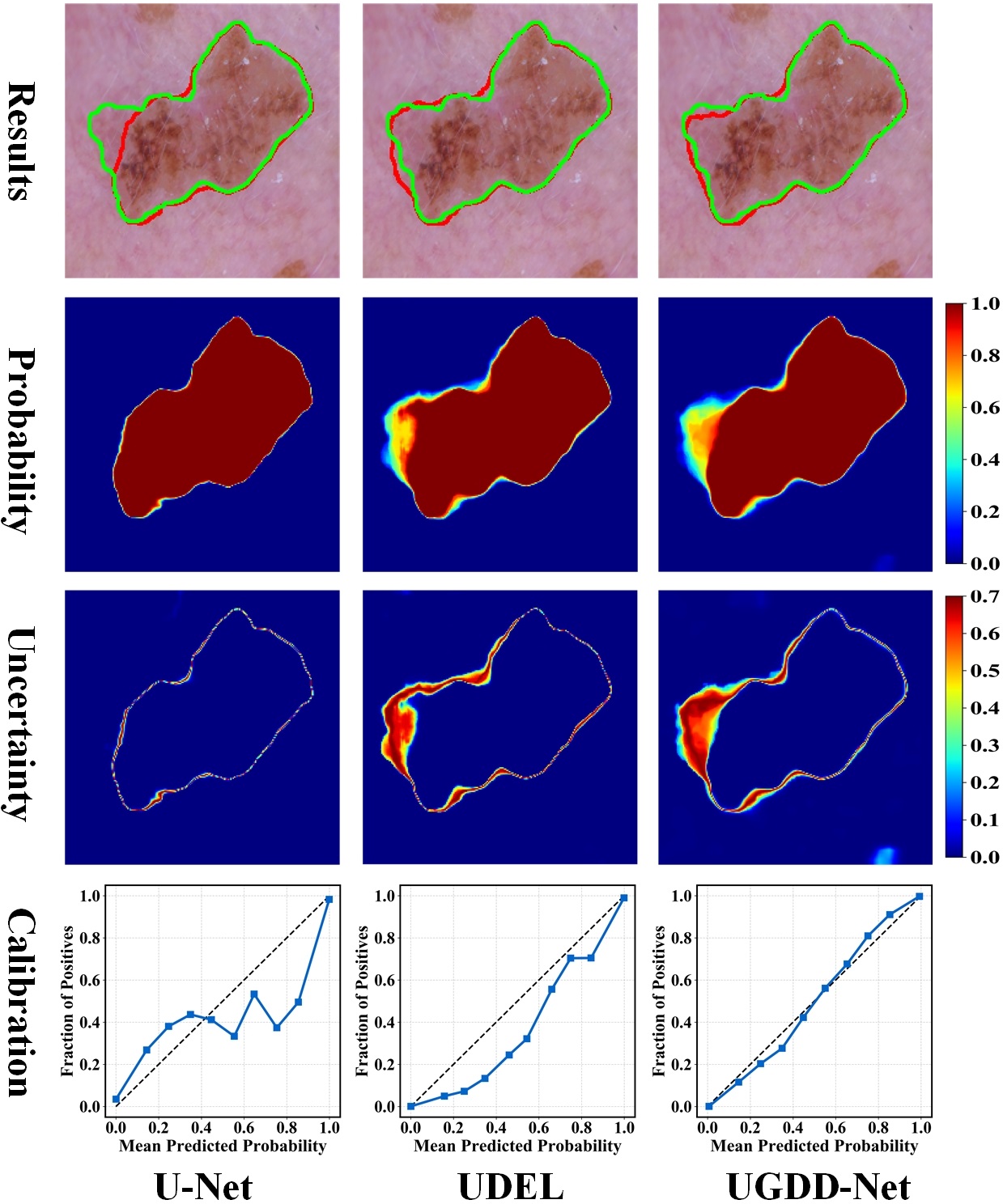}
	\caption{Qualitative comparison of the vanilla U-Net, the recent uncertainty-aware method UDEL \cite{wang2025udel}, and our proposed UGDD-Net on a challenging skin lesion. 
		[Row 1 (Results)]: Segmentation contours (Green: Ground truth, Red: Prediction). UGDD-Net achieves the most precise boundary delineation. 
		[Row 2 (Probability)]: The dense probability heatmaps before binarization. 
		[Row 3 (Uncertainty)]: Estimated uncertainty maps, derived by calculating the pixel-wise entropy of the probability distributions in Row 2. Vanilla U-Net exhibits blind certainty, yielding only thin, uninformative lines. While UDEL captures broader uncertainty, it aligns poorly with the actual visual ambiguity of the original image, seemingly just reacting to its own segmentation errors rather than identifying true diagnostic difficulty. In contrast, our uncertainty precisely highlights the intrinsically blurred regions of the lesion, providing diagnostically meaningful insights to guide clinical attention. 
		[Row 4 (Calibration)]: Reliability diagrams. Unlike U-Net's erratic fluctuations and UDEL's severe over-confidence (falling significantly below the diagonal), our UGDD-Net tightly hugs the ideal reference line, demonstrating superior statistical calibration. This confirms that the model's predicted confidence can be relied upon by dermatologists to assess diagnostic risks.}
	\label{fig:motivation}
\end{figure}

With the advent of deep learning, convolutional networks, represented by U-Net \cite{ronneberger2015u} and its variants \cite{zhou2022h, dai2025improving}, have established the dominant paradigm for skin lesion segmentation. To transcend the localized receptive fields of CNNs, recent architectures have incorporated Vision Transformers (ViTs) \cite{cao2022swin, ji2025bdformer} and State Space Models (e.g., Mamba) \cite{ruan2024vm, liu2025vision} to capture long-range global dependencies. Despite these architectural strides, a fundamental bottleneck persists: the majority of existing models operate exclusively in the spatial domain. However, pathologically, the disorganized infiltration of abnormal cells and the uneven deposition of melanin frequently result in diffuse transitions lacking sharp anatomical borders. Consequently, these boundaries exhibit vanishingly weak spatial gradients, rendering spatial pixel intensity alone insufficient to precisely separate the lesion from surrounding healthy skin. To address this issue, recent research has pivoted towards multi-representation context modeling. For instance, some studies have introduced frequency-domain methods to isolate high-frequency boundary details \cite{liu2025sfma, cai2026cdmt}, while others employ Graph Neural Networks (GNNs) to reason about irregular spatial topologies \cite{rahman2024g, kui2025wingraphunet}. While these multi-representation approaches theoretically offer a more comprehensive description than spatial features alone, they inadvertently introduce a new set of critical vulnerabilities.

We contend that these vulnerabilities stem from a systemic oversight: the inadequate active perception and utilization of uncertainty to guide multi-representation learning. Even when quantified, uncertainty is typically relegated to a passive, post-hoc evaluation metric \cite{zou2025toward, munia2025attention} rather than an active learning prior. This oversight traps models in a deterministic framework, manifesting in three critical bottlenecks: \textbf{(\textit{i}) Blind Fusion:} Conventional networks typically merge cross-domain features via straightforward concatenation or uncertainty-blind attention \cite{zhou2024spatial, huang2025dpmf}. Lacking explicit reliability metrics to decouple feature salience from diagnostic trustworthiness, these reliability-agnostic operations fail to effectively filter domain-specific weaknesses. Consequently, high-frequency spectral artifacts and spatial ambiguities propagate bidirectionally, corrupting the joint representation. \textbf{(\textit{ii}) Passive Reasoning:} Beyond initial feature fusion, the subsequent structural reasoning process also suffers from this deterministic bias. While some methods incorporate topological modeling (e.g., GNNs), they typically construct static graphs based on raw feature similarity rather than prediction confidence \cite{kui2025wingraphunet, li2026learning}. Lacking an uncertainty-aware propagation mechanism, they cannot leverage high-confidence semantic consensus to actively refine low-confidence nodes, often resulting in foreground-background confusion and distorted lesion boundaries. \textbf{(\textit{iii}) Rigid Optimization:} Finally, the potential of these rich features is severely stifled at the optimization end. Conventional loss functions (e.g., Cross-Entropy \cite{shore2003properties}, Dice \cite{milletari2016v}), dynamically weighted variants like Focal Loss \cite{lin2017focal}, and even explicit boundary constraints \cite{kervadec2019boundary} impose uncompromising penalties based on static labels. By aggressively penalizing inherently ambiguous pixels, this rigid supervision forces the network to overfit subjective label noise, yielding severely over-confident predictions and statistical miscalibration that render the model clinically unreliable (as evidenced in Fig. \ref{fig:motivation}).

To dismantle these bottlenecks, we propose the Uncertainty-Guided Dual-Domain Network (UGDD-Net). By synergizing spatial and frequency-domain representations within an uncertainty-in-the-loop unified architecture, our framework enables end-to-end reliability propagation and optimization. 

Specifically, the contributions of this work are as follows:

\begin{itemize}
	\item \textbf{Uncertainty-Guided Dual-Domain Network (UGDD-Net):} We propose a closed-loop framework that transforms uncertainty from a passive posterior metric into an active guiding signal that permeates the entire learning lifecycle. Powered by the novel two-pass ``glance-and-gaze'' forward mechanism, UGDD-Net decouples the segmentation process into an initial global ``glance'' and a subsequent uncertainty-driven ``gaze.'' This effectively breaks the circular dependency of uncertainty estimation, enabling the model to self-diagnose and self-rectify across parallel spatial and spectral pathways.
	
	\item \textbf{Uncertainty-Guided Bi-directional Feature Fusion (UGBFF):} To resolve the ``blind fusion'' dilemma, UGBFF employs pixel-level uncertainty to dynamically modulate cross-domain interactions. By allowing high-confidence representations from one domain to actively supervise and rectify unreliable features in the other, it ensures reliability-modulated feature integration while facilitating reciprocal enhancement.
	
	\item \textbf{Uncertainty-Guided Graph Refinement (UGGR):} To overcome ``passive reasoning,'' UGGR constructs a sparse graph for topology-aware relational inference. By utilizing a priority-guided cross-attention mechanism, it actively propagates reliable semantic consensus to rectify uncertain query nodes, effectively ensuring global semantic and morphological consistency.
	
	\item \textbf{Uncertainty-Guided Margin-Adaptive Loss (UGML):} To address ``rigid optimization,'' UGML dynamically adapts decision margins based on pixel-wise uncertainty. By enforcing strict constraints on confident pixels while relaxing penalties on uncertain ones, it ensures the model is decisively confident in clear regions yet appropriately cautious in ambiguous ones, effectively suppressing erratic confidence fluctuations to achieve superior statistical calibration.
	
	\item \textbf{Comprehensive Validation and Clinical Interpretability:} Extensive experiments on four public datasets (ISIC2017, ISIC2018, PH$^2$, and HAM10000) demonstrate that UGDD-Net establishes a new state-of-the-art, with its superiority particularly pronounced on our curated ``Hard Samples'' subset. Moreover, our uncertainty maps exhibit meaningful alignment with expert inter-observer variability on diagnostically ambiguous images, providing robust interpretability for human-machine collaborative diagnosis.
\end{itemize}

\section{Results}\label{sec2}

\subsection{Datasets and Evaluation Protocols}
\label{Datasets}
To comprehensively evaluate the performance and robustness of the proposed UGDD-Net, we utilized four public skin lesion segmentation datasets: ISIC2017 \cite{codella2018skin}, ISIC2018 \cite{bissoto2018deep}, $\text{PH}^2$ \cite{mendoncca2013ph}, and HAM10000 \cite{tschandl2018ham10000}. 

\subsubsection{Public ISIC Benchmarks: ISIC2017 \textup{\&} ISIC2018}
We benchmarked our method on the two most widely used datasets in the field.
\begin{itemize}
	\item \textbf{ISIC2017 Dataset:} Comprises 2,000 training, 150 validation, and 600 test images. We adhered to the official fixed splits. Results are averaged over five independent runs to ensure reliability.
	\item \textbf{ISIC2018 Dataset:} Contains 2,594 images. As the official test set ground truth remains unreleased, we employed a 5-fold cross-validation strategy (70\% training, 10\% validation, 20\% testing). Final results represent the average across the five folds.
\end{itemize}

\subsubsection{Generalization Assessment: Cross-Dataset Protocol}
To evaluate generalization capability on unseen data distributions, we established a bidirectional cross-dataset protocol using the $\text{PH}^2$ dataset (200 images) and ISIC2018:
\begin{itemize}
	\item \textbf{ISIC2018 $\rightarrow$ $\text{PH}^2$:} The ISIC2018 dataset was randomly split into training (80\%) and validation (20\%) sets for model optimization, and the trained model was directly tested on the entire $\text{PH}^2$ dataset.
	\item \textbf{$\text{PH}^2$ $\rightarrow$ ISIC2018:} Symmetrically, the $\text{PH}^2$ dataset was randomly partitioned into training (80\%) and validation (20\%) sets, while the entire ISIC2018 dataset was used strictly for testing.
\end{itemize}

We repeated these cross-dataset experiments five times with different random seeds and reported the average performance to ensure statistical reliability.

\subsubsection{Robustness Evaluation: Hard Sample Protocol on HAM10000}
Standard holistic metrics often mask performance degradation on specific failure modes due to the dominance of ``easy'' samples. To rigorously stress-test the model's robustness, we introduced a hard sample evaluation protocol using the HAM10000 dataset (10,015 images).

\textbf{Hard Sample Selection Strategy:} To objectively identify challenging cases, we used a standard U-Net \cite{ronneberger2015u} as a baseline probe. Through 5-fold cross-validation on the entire HAM10000 dataset, we calculated U-Net's global average IoU (${\sim}86.75\%$). Samples with a predicted IoU below this threshold were categorized as Hard Samples. These images typically feature small lesions, ambiguous boundaries, irregular topologies, or significant artifacts.

Based on this strategy, we evaluated our model on two distinct subsets:
\begin{itemize}
	\item \textbf{All Samples:} The complete dataset, providing a holistic view of general segmentation capability.
	\item \textbf{Hard Samples:} The filtered challenging cases. This subset serves as a critical stress test to evaluate robustness in complex scenarios where baseline architectures struggle.
\end{itemize}

\subsection{Evaluation Metrics and Implementation Details}

\subsubsection{Evaluation Metrics}
We comprehensively evaluated the proposed method using five standard metrics grouped into three functional categories:

\textbf{Area-Based Metrics:} Intersection over Union (IoU) and the Dice Similarity Coefficient (Dice) quantify the regional overlap between predictions and ground truth, where higher scores indicate better segmentation accuracy.

\textbf{Boundary-Based Metrics:} To evaluate contour precision, we employed the 95th percentile Hausdorff Distance (HD95) to robustly assess the worst-case mismatch, and the Average Symmetric Surface Distance (ASSD) to reflect global shape consistency. For these metrics, lower values denote superior performance.

\textbf{Reliability Metric:} The Expected Calibration Error (ECE) evaluates the trustworthiness of the model's uncertainty estimation. It measures the expected gap between predicted confidence and actual accuracy; a lower ECE indicates a better-calibrated model whose predicted probabilities closely align with true correctness.

\textbf{Statistical Analysis:} To assess the significance of our improvements, we conducted a paired $t$-test comparing UGDD-Net against the second-best method across all metrics. Following standard statistical practices \cite{altman2017points}, $p < 0.05$ and $p < 0.01$ are used to denote statistically significant and highly statistically significant differences, respectively.

\subsection{Performance Comparison on ISIC Benchmarks}

\subsubsection{Quantitative Analysis}

\textbf{Overall Superiority:} As shown in Table \ref{tab:isic_comparison}, UGDD-Net outperforms the second-best method (UDEL, underlined) by a highly statistically significant margin ($p < 0.01$) across the board. Specifically, on the ISIC2017 dataset, our method improves the IoU score by 2.40\% (81.97\% vs. 79.57\%) and reduces the HD95 by 1.93 pixels (12.68 vs. 14.61). On the ISIC2018 dataset, UGDD-Net achieves an outstanding Dice score of 91.26\% (a 1.42\% improvement) and reduces the ASSD from 4.78 to a mere 3.53 pixels. These balanced gains highlight the effectiveness of our dual-domain architecture in handling complex morphological variations and boundary ambiguities.

\begin{table*}[!ht]
	\centering
    \caption{Quantitative comparison of skin lesion segmentation performance against state-of-the-art methods on the ISIC2017 and ISIC2018 datasets. {\sfmark} denotes spatial-frequency methods, and {\uncmark} denotes uncertainty-aware methods.}
	\label{tab:isic_comparison}
	
	\newcommand{\spm}{\!\pm\mkern 2mu\scriptstyle} 
	
	\resizebox{\textwidth}{!}{%
		\begin{threeparttable}
			\begin{tabular}{lccccccccccc}
				\toprule
				\multirow{2}{*}{\textbf{Model}} & \multirow{2}{*}{\textbf{Backbone}} & \multicolumn{5}{c}{\textbf{ISIC2017}} & \multicolumn{5}{c}{\textbf{ISIC2018}} \\ \cmidrule(lr){3-7} \cmidrule(lr){8-12}
				& & IoU \scriptsize{(\%)} $\uparrow$ & Dice \scriptsize{(\%)} $\uparrow$ & HD95 \scriptsize{(px)} $\downarrow$ & ASSD \scriptsize{(px)} $\downarrow$ & ECE \scriptsize{(\%)} $\downarrow$ & IoU \scriptsize{(\%)} $\uparrow$ & Dice \scriptsize{(\%)} $\uparrow$ & HD95 \scriptsize{(px)} $\downarrow$ & ASSD \scriptsize{(px)} $\downarrow$ & ECE \scriptsize{(\%)} $\downarrow$ \\
				\midrule
				\multicolumn{12}{l}{\textit{Single-Path Methods}} \\
				U-Net \cite{ronneberger2015u} & CNN & $74.94\spm0.96$ & $83.87\spm0.80$ & $19.11\spm1.19$ & $7.42\spm0.99$ & $15.77\spm1.51$ & $80.21\spm1.12$ & $87.77\spm0.87$ & $15.51\spm1.34$ & $5.88\spm1.14$ & $14.75\spm1.66$ \\
				Swin-Unet \cite{cao2022swin} & Trans & $74.63\spm1.20$ & $83.46\spm1.01$ & $19.85\spm1.50$ & $7.68\spm1.30$ & $19.79\spm1.90$ & $80.73\spm1.40$ & $88.09\spm1.10$ & $16.28\spm1.70$ & $6.06\spm1.50$ & $17.44\spm2.10$ \\
				VM-UNet \cite{ruan2024vm} & Mamba & $76.38\spm0.70$ & $84.87\spm0.57$ & $18.92\spm0.84$ & $7.31\spm0.66$ & $16.26\spm1.07$ & $80.79\spm0.81$ & $88.15\spm0.62$ & $14.48\spm0.94$ & $5.77\spm0.75$ & $15.43\spm1.17$ \\
				MPBA-Net \cite{huang2025lesion} & CNN+Trans & $78.11\spm0.59$ & $86.17\spm0.46$ & $16.93\spm0.68$ & $6.82\spm0.51$ & $17.57\spm0.87$ & $81.37\spm0.67$ & $88.70\spm0.51$ & $12.88\spm0.75$ & $5.33\spm0.58$ & $16.21\spm0.94$ \\
				WinGraphUNet \cite{kui2025wingraphunet} & CNN+Graph & $77.94\spm0.54$ & $86.10\spm0.42$ & $14.74\spm0.62$ & $6.02\spm0.44$ & $17.84\spm0.80$ & $81.42\spm0.61$ & $88.69\spm0.46$ & $11.32\spm0.68$ & $4.76\spm0.50$ & $16.22\spm0.86$ \\ 
				CDMT-UNet\sfmark \cite{cai2026cdmt} & CNN+Trans & $78.22\spm0.92$ & $86.26\spm0.76$ & $15.65\spm1.12$ & $6.32\spm0.93$ & $18.65\spm1.43$ & $82.83\spm1.06$ & $89.80\spm0.83$ & $11.83\spm1.26$ & $4.96\spm1.07$ & $16.71\spm1.57$ \\ 				
				SkinMamba\sfmark \cite{zou2024skinmamba} & CNN+Mamba & $76.88\spm0.68$ & $85.26\spm0.55$ & $15.63\spm0.81$ & $6.37\spm0.63$ & $16.27\spm1.03$ & $82.66\spm0.78$ & $89.64\spm0.60$ & $11.46\spm0.90$ & $4.87\spm0.72$ & $15.86\spm1.12$ \\
				DEviS\uncmark \cite{zou2025toward} & CNN & $77.16\spm0.35$ & $85.43\spm0.25$ & $17.45\spm0.37$ & $7.15\spm0.20$ & $13.91\spm0.48$ & $81.92\spm0.39$ & $88.96\spm0.28$ & $13.42\spm0.39$ & $5.37\spm0.22$ & $12.75\spm0.50$ \\
				AHF-U-Net\uncmark \cite{munia2025attention} & CNN & $77.02\spm0.63$ & $85.38\spm0.50$ & $17.29\spm0.75$ & $6.91\spm0.57$ & $14.76\spm0.95$ & $81.71\spm0.73$ & $88.84\spm0.55$ & $13.72\spm0.83$ & $5.47\spm0.65$ & $13.89\spm1.03$ \\
				DPGNet\uncmark \cite{wang2025dpgnet} & CNN+Trans & $77.34\spm0.44$ & $85.68\spm0.33$ & $16.98\spm0.50$ & $6.86\spm0.32$ & $14.38\spm0.64$ & $81.98\spm0.50$ & $89.11\spm0.37$ & $13.45\spm0.54$ & $5.41\spm0.36$ & $13.52\spm0.68$ \\				
				\midrule 
				\multicolumn{12}{l}{\textit{Dual-Path Methods}} \\
				H-Net \cite{zhou2022h} & CNN & $75.98\spm0.87$ & $84.54\spm0.71$ & $19.03\spm1.06$ & $7.35\spm0.87$ & $15.87\spm1.35$ & $80.44\spm1.01$ & $88.01\spm0.78$ & $14.89\spm1.19$ & $5.85\spm1.00$ & $15.28\spm1.48$ \\
				BDFormer \cite{ji2025bdformer} & Trans & $76.42\spm1.15$ & $84.85\spm0.97$ & $16.38\spm1.44$ & $6.69\spm1.24$ & $19.44\spm1.82$ & $81.83\spm1.34$ & $88.91\spm1.05$ & $12.56\spm1.63$ & $5.25\spm1.43$ & $17.25\spm2.01$ \\
				CFFormer \cite{li2026cfformer} & CNN+Trans & $76.97\spm1.11$ & $85.32\spm0.93$ & $17.69\spm1.37$ & $7.20\spm1.18$ & $18.15\spm1.74$ & $82.71\spm1.29$ & $89.72\spm1.01$ & $13.84\spm1.55$ & $5.48\spm1.36$ & $16.57\spm1.92$ \\
				DSU-Net \cite{zhong2025dsu} & CNN+Trans & $76.90\spm1.01$ & $85.25\spm0.84$ & $18.59\spm1.25$ & $7.24\spm1.06$ & $18.04\spm1.58$ & $81.60\spm1.18$ & $88.74\spm0.92$ & $14.27\spm1.41$ & $5.70\spm1.22$ & $16.58\spm1.74$ \\
				DMA-Net \cite{zhai2024dma} & CNN+Mamba & $76.74\spm1.06$ & $85.06\spm0.88$ & $17.57\spm1.31$ & $7.16\spm1.12$ & $16.38\spm1.66$ & $81.02\spm1.23$ & $88.41\spm0.96$ & $13.87\spm1.48$ & $5.55\spm1.29$ & $15.93\spm1.83$ \\
				SF-UNet\sfmark \cite{zhou2024spatial} & CNN & $76.64\spm0.82$ & $84.95\spm0.67$ & $16.07\spm1.00$ & $6.48\spm0.81$ & $15.94\spm1.27$ & $80.94\spm0.95$ & $88.23\spm0.74$ & $12.39\spm1.12$ & $5.08\spm0.93$ & $14.96\spm1.39$ \\
				SFma-UNet\sfmark \cite{liu2025sfma} & Mamba & $78.99\spm0.73$ & $86.91\spm0.59$ & $15.38\spm0.87$ & $6.25\spm0.69$ & $16.92\spm1.11$ & $82.86\spm0.84$ & $89.75\spm0.64$ & $11.40\spm0.97$ & $4.84\spm0.79$ & $16.16\spm1.21$ \\				
				DPMF-Net\sfmark \cite{huang2025dpmf} & CNN+Trans & $77.85\spm0.78$ & $85.96\spm0.63$ & $15.84\spm0.94$ & $6.45\spm0.75$ & $16.14\spm1.19$ & $82.47\spm0.90$ & $89.56\spm0.69$ & $11.73\spm1.05$ & $5.01\spm0.86$ & $15.33\spm1.30$ \\
				F2CAU-Net\uncmark \cite{zhou2025f2cau} & CNN & $78.15\spm0.49$ & $86.16\spm0.38$ & $16.28\spm0.56$ & $6.62\spm0.38$ & $14.11\spm0.72$ & $82.34\spm0.56$ & $89.32\spm0.42$ & $12.48\spm0.61$ & $5.15\spm0.43$ & $13.17\spm0.77$ \\				
				UDEL\uncmark \cite{wang2025udel} & CNN+Mamba & \underline{$79.57\spm0.40$} & \underline{$87.34\spm0.29$} & \underline{$14.61\spm0.43$} & \underline{$5.96\spm0.26$} & \underline{$13.47\spm0.56$} & \underline{$83.12\spm0.45$} & \underline{$89.84\spm0.33$} & \underline{$11.32\spm0.46$} & \underline{$4.78\spm0.29$} & \underline{$12.49\spm0.59$} \\      
				\midrule 
				\textbf{UGDD-Net (Ours)}\bothmark & Dual-Domain & $\mathbf{81.97}\spm\mathbf{0.32}$\rlap{$^{*}$} & $\mathbf{89.35}\spm\mathbf{0.21}$\rlap{$^{*}$} & $\mathbf{12.68}\spm\mathbf{0.33}$\rlap{$^{*}$} & $\mathbf{4.72}\spm\mathbf{0.16}$\rlap{$^{*}$} & $\mathbf{10.92}\spm\mathbf{0.42}$\rlap{$^{*}$} & $\mathbf{84.69}\spm\mathbf{0.36}$\rlap{$^{*}$} & $\mathbf{91.26}\spm\mathbf{0.24}$\rlap{$^{*}$} & $\mathbf{9.86}\spm\mathbf{0.35}$\rlap{$^{*}$} & $\mathbf{3.53}\spm\mathbf{0.17}$\rlap{$^{*}$} & $\mathbf{10.03}\spm\mathbf{0.44}$\rlap{$^{*}$} \\
				\bottomrule
			\end{tabular}
			\begin{tablenotes}
				\item \textbf{Note:} \textbf{Bold} and \underline{underline} denote the best and second-best results, respectively. $^{*}$ indicates a highly statistically significant improvement ($p < 0.01$) vs. the second-best method.
			\end{tablenotes}
		\end{threeparttable}%
	}
\end{table*}

\textbf{Superiority Over Spatial-Frequency and Uncertainty Methods:} A critical distinction emerges when comparing against methods employing similar sub-mechanisms:
\begin{itemize}
	\item \textbf{Spatial-Frequency Methods} ($^{\textcolor{blue}{\mathcal{SF}}}$): Existing networks such as SFma-UNet and CDMT-UNet successfully extract both spatial semantics and high-frequency details. However, their deterministic cross-fusion mechanisms lack active reliability perception. Consequently, this ``blind fusion'' allows high-frequency artifacts (e.g., hair, markers) to propagate bidirectionally. In contrast, our Uncertainty-Guided Bi-Directional Feature Fusion (UGBFF) acts as an active filter, utilizing uncertainty to dynamically suppress these artifact responses, yielding significantly better boundary precision.
	
	\item \textbf{Uncertainty-Aware Methods} ($^{\textcolor{red}{\mathcal{U}}}$): Methods such as UDEL and DPGNet explicitly estimate uncertainty to perceive boundaries or dynamically adjust learning weights. However, their utilization of uncertainty remains limited, largely confined to spatial-domain attention or loss re-weighting. UGDD-Net extends uncertainty from a localized metric to an active prior that permeates the entire framework. By tightly coupling uncertainty with dual-domain fusion (UGBFF) and global graph reasoning (UGGR), our model can explicitly resolve conflicting signals between artifacts and true boundaries, leading to superior robustness.
\end{itemize}

\subsubsection{Qualitative Visualization}
\begin{figure*}[!ht]
	\centering
	\includegraphics[width=0.95\linewidth]{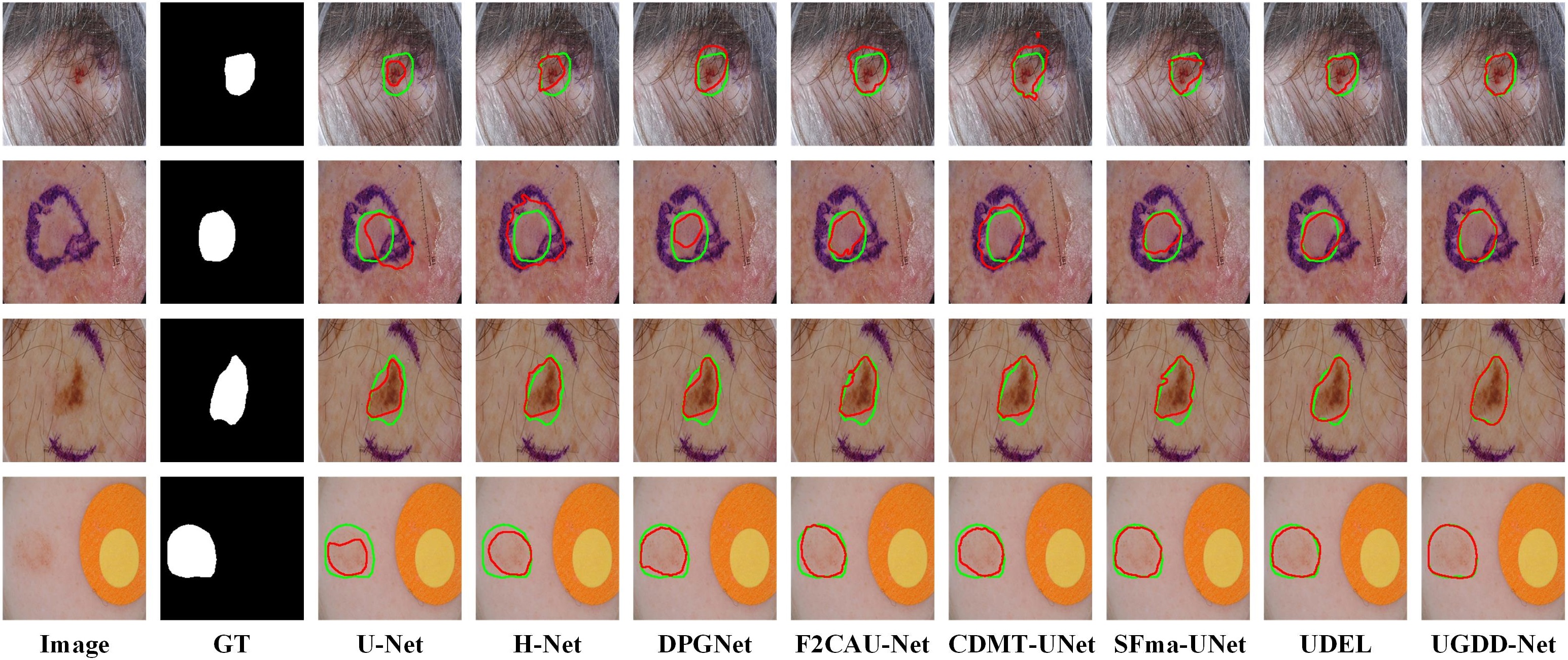}
	\caption{Qualitative visualization of skin lesion segmentation on the ISIC2017 dataset (first two rows) and the ISIC2018 dataset (last two rows). The selected samples present severe challenges, including dense hair, prominent markers, and blurred boundaries. The green curves outline the ground truth contours, and the red curves indicate the predicted boundaries.}
	\label{fig:isic_visual}
\end{figure*}

To intuitively evaluate the segmentation quality, we visualize representative results in Fig. \ref{fig:isic_visual} (the first two rows from ISIC2017, and the last two from ISIC2018). Notably, all selected samples present varying degrees of artifact interference coupled with morphological challenges.

The lesion in the first row is heavily obscured by dense hair, while the second is severely interfered with by a prominent blue marker circle. Baseline models (e.g., U-Net, H-Net) and even some frequency-aware methods suffer from ``passive perception,'' yielding severely distorted contours or mistakenly including these artifacts into the foreground. Driven by the UGBFF module, UGDD-Net explicitly assigns high uncertainty to these artifact regions and actively filters them out, thereby preventing severe contour distortion.

The third row exhibits an irregular, blurred lesion disturbed by a blue marker, whereas the fourth row shows a blurred lesion accompanied by color chart patches. Without a dynamic mechanism to distinguish true weak edges from strong surrounding noise, conventional deterministic models tend to either over-segment the noise or under-segment the fuzzy boundaries. Guided by the frequency branch and the dynamic margin loss (UGML), our method robustly disentangles the lesions from the complex backgrounds, generating precise contours that closely align with the ground truth.

\subsection{Cross-Dataset Generalization Analysis}

\subsubsection{Quantitative Analysis}
Generalization across different clinical centers remains a critical bottleneck for deep learning models due to severe domain shifts in imaging devices, illumination, and clinical protocols. Table~\ref{tab:cross_dataset} presents the cross-dataset evaluation results. UGDD-Net demonstrates remarkable robustness, outperforming all competing methods by a highly statistically significant margin ($p < 0.01$) in both generalization directions.

\begin{table*}[!ht]
	\centering
	\caption{Quantitative comparison of cross-dataset generalization performance. The models are trained on the ISIC2018 dataset and tested on the PH$^2$ dataset (ISIC2018 $\to$ PH$^2$), and vice versa (PH$^2$ $\to$ ISIC2018). {\sfmark} denotes Spatial-Frequency methods, and {\uncmark} denotes Uncertainty-aware methods.}
	\label{tab:cross_dataset}
	
	\newcommand{\spm}{\!\pm\mkern 2mu\scriptstyle} 
	
	\resizebox{\textwidth}{!}{%
		\begin{threeparttable}
			\begin{tabular}{lcccccccccc}
				\toprule
				\multirow{2}{*}{\textbf{Model}} & \multicolumn{5}{c}{\textbf{ISIC2018 $\to$ PH$^2$}} & \multicolumn{5}{c}{\textbf{PH$^2$ $\to$ ISIC2018}} \\
				\cmidrule(lr){2-6} \cmidrule(lr){7-11}
				& IoU \scriptsize{(\%)} $\uparrow$ & Dice \scriptsize{(\%)} $\uparrow$ & HD95 \scriptsize{(px)} $\downarrow$ & ASSD \scriptsize{(px)} $\downarrow$ & ECE \scriptsize{(\%)} $\downarrow$ & IoU \scriptsize{(\%)} $\uparrow$ & Dice \scriptsize{(\%)} $\uparrow$ & HD95 \scriptsize{(px)} $\downarrow$ & ASSD \scriptsize{(px)} $\downarrow$ & ECE \scriptsize{(\%)} $\downarrow$ \\
				\midrule
				U-Net \citep{ronneberger2015u}             & $79.81\spm1.00$ & $87.95\spm0.84$ & $17.15\spm1.25$ & $6.89\spm1.05$ & $15.87\spm1.57$ & $64.59\spm1.17$ & $74.95\spm0.92$ & $25.52\spm1.40$ & $10.94\spm1.21$ & $16.68\spm1.77$ \\
				Swin-Unet \citep{cao2022swin}            & $79.93\spm1.22$ & $88.14\spm1.05$ & $18.46\spm1.55$ & $7.31\spm1.35$ & $20.47\spm1.95$ & $63.75\spm1.45$ & $74.23\spm1.15$ & $25.91\spm1.75$ & $11.15\spm1.55$ & $20.93\spm2.20$ \\
				VM-UNet \citep{ruan2024vm}                & $80.69\spm0.73$ & $88.45\spm0.59$ & $17.03\spm0.89$ & $6.92\spm0.70$ & $16.74\spm1.12$ & $64.21\spm0.84$ & $74.56\spm0.65$ & $25.47\spm0.98$ & $10.81\spm0.80$ & $17.28\spm1.24$ \\
				MPBA-Net \citep{huang2025lesion}          & $81.91\spm0.59$ & $89.39\spm0.47$ & $13.56\spm0.71$ & $5.92\spm0.52$ & $18.48\spm0.90$ & $67.84\spm0.68$ & $78.72\spm0.52$ & $23.90\spm0.77$ & $9.58\spm0.60$ & $18.66\spm0.98$ \\
				WinGraphUNet \citep{kui2025wingraphunet}  & $82.42\spm0.55$ & $89.63\spm0.43$ & $12.41\spm0.65$ & $5.35\spm0.46$ & $19.27\spm0.82$ & $65.85\spm0.62$ & $76.13\spm0.47$ & $22.18\spm0.70$ & $8.07\spm0.53$ & $18.74\spm0.90$ \\ 
				CDMT-UNet\sfmark \citep{cai2026cdmt}        & $83.93\spm0.91$ & $90.82\spm0.76$ & $12.45\spm1.13$ & $5.44\spm0.93$ & $19.89\spm1.42$ & $70.79\spm1.06$ & $80.46\spm0.83$ & $22.49\spm1.26$ & $8.52\spm1.07$ & $20.32\spm1.59$ \\ 
				SkinMamba\sfmark \citep{zou2024skinmamba}   & $82.87\spm0.68$ & $90.29\spm0.55$ & $12.73\spm0.83$ & $5.52\spm0.64$ & $17.32\spm1.05$ & $66.25\spm0.79$ & $76.67\spm0.61$ & $22.36\spm0.91$ & $8.23\spm0.74$ & $17.32\spm1.16$ \\
				DEviS\uncmark \citep{zou2025toward}            & $83.15\spm0.37$ & $90.36\spm0.26$ & $14.07\spm0.41$ & $6.22\spm0.22$ & $14.17\spm0.52$ & $67.55\spm0.40$ & $77.56\spm0.29$ & $24.81\spm0.42$ & $10.32\spm0.26$ & $14.86\spm0.55$ \\
				AHF-U-Net\uncmark \citep{munia2025attention}  & $81.74\spm0.64$ & $89.30\spm0.51$ & $14.08\spm0.77$ & $6.15\spm0.58$ & $14.60\spm0.97$ & $66.86\spm0.73$ & $77.04\spm0.56$ & $24.58\spm0.84$ & $10.02\spm0.67$ & $15.95\spm1.07$ \\
				DPGNet\uncmark \citep{wang2025dpgnet}         & $83.01\spm0.46$ & $90.12\spm0.34$ & $13.84\spm0.53$ & $6.09\spm0.34$ & $14.36\spm0.67$ & $70.44\spm0.51$ & $80.73\spm0.38$ & $23.28\spm0.56$ & $9.17\spm0.40$ & $15.73\spm0.72$ \\
				\midrule 
				H-Net \citep{zhou2022h}                  & $79.67\spm0.86$ & $87.73\spm0.72$ & $17.52\spm1.07$ & $7.06\spm0.87$ & $15.45\spm1.35$ & $64.48\spm1.01$ & $74.72\spm0.79$ & $25.69\spm1.19$ & $11.04\spm1.01$ & $16.79\spm1.51$ \\
				BDFormer \citep{ji2025bdformer}          & $82.56\spm1.17$ & $89.89\spm1.01$ & $13.44\spm1.49$ & $5.72\spm1.29$ & $20.29\spm1.88$ & $65.03\spm1.40$ & $75.58\spm1.11$ & $23.75\spm1.68$ & $9.43\spm1.48$ & $20.89\spm2.11$ \\
				CFFormer \citep{li2026cfformer}          & $83.76\spm1.09$ & $90.71\spm0.93$ & $14.95\spm1.37$ & $6.30\spm1.17$ & $19.73\spm1.72$ & $69.24\spm1.28$ & $79.38\spm1.01$ & $25.06\spm1.54$ & $10.52\spm1.35$ & $20.32\spm1.94$ \\
				DSU-Net \citep{zhong2025dsu}             & $82.54\spm1.04$ & $89.77\spm0.88$ & $16.63\spm1.31$ & $6.77\spm1.11$ & $19.28\spm1.65$ & $65.37\spm1.23$ & $75.63\spm0.97$ & $25.24\spm1.47$ & $10.58\spm1.28$ & $19.72\spm1.85$ \\
				DMA-Net \citep{zhai2024dma}              & $81.41\spm1.09$ & $89.19\spm0.93$ & $14.60\spm1.37$ & $6.26\spm1.17$ & $17.36\spm1.72$ & $65.74\spm1.28$ & $75.79\spm1.01$ & $24.86\spm1.54$ & $10.43\spm1.35$ & $17.90\spm1.94$ \\
				SF-UNet\sfmark \citep{zhou2024spatial}      & $80.84\spm0.86$ & $88.58\spm0.72$ & $13.33\spm1.07$ & $5.74\spm0.87$ & $15.64\spm1.35$ & $68.72\spm1.01$ & $77.97\spm0.79$ & $23.12\spm1.19$ & $8.91\spm1.01$ & $16.89\spm1.51$ \\
				SFma-UNet\sfmark \citep{liu2025sfma}       & $83.88\spm0.77$ & $90.79\spm0.63$ & $12.65\spm0.95$ & $5.43\spm0.76$ & $17.67\spm1.20$ & $70.89\spm0.90$ & $81.11\spm0.70$ & $22.21\spm1.05$ & $8.15\spm0.87$ & $18.03\spm1.33$ \\
				DPMF-Net\sfmark \citep{huang2025dpmf}      & $83.23\spm0.82$ & $90.40\spm0.68$ & $13.02\spm1.01$ & $5.63\spm0.81$ & $16.52\spm1.27$ & $70.18\spm0.95$ & $79.94\spm0.74$ & $22.98\spm1.12$ & $8.72\spm0.94$ & $17.20\spm1.42$ \\
				F2CAU-Net\uncmark \citep{zhou2025f2cau}      & $83.19\spm0.50$ & $90.33\spm0.38$ & $13.55\spm0.59$ & $5.75\spm0.40$ & $14.25\spm0.75$ & $69.01\spm0.57$ & $79.05\spm0.43$ & $24.26\spm0.63$ & $9.84\spm0.46$ & $15.38\spm0.81$ \\
				UDEL\uncmark \citep{wang2025udel}            & \underline{$84.01\spm0.41$} & \underline{$90.96\spm0.30$} & \underline{$12.25\spm0.47$} & \underline{$5.23\spm0.28$} & \underline{$13.76\spm0.60$} & \underline{$71.27\spm0.46$} & \underline{$81.36\spm0.34$} & \underline{$22.02\spm0.49$} & \underline{$7.95\spm0.33$} & \underline{$14.59\spm0.64$} \\      
				\midrule 
				\textbf{UGDD-Net (Ours)}\bothmark            & $\mathbf{85.93}\spm\mathbf{0.34}$\rlap{$^{*}$} & $\mathbf{92.07}\spm\mathbf{0.22}$\rlap{$^{*}$} & $\mathbf{11.07}\spm\mathbf{0.37}$\rlap{$^{*}$} & $\mathbf{4.26}\spm\mathbf{0.18}$\rlap{$^{*}$} & $\mathbf{11.24}\spm\mathbf{0.46}$\rlap{$^{*}$} & $\mathbf{73.52}\spm\mathbf{0.37}$\rlap{$^{*}$} & $\mathbf{83.29}\spm\mathbf{0.25}$\rlap{$^{*}$} & $\mathbf{20.23}\spm\mathbf{0.38}$\rlap{$^{*}$} & $\mathbf{6.64}\spm\mathbf{0.21}$\rlap{$^{*}$} & $\mathbf{12.02}\spm\mathbf{0.49}$\rlap{$^{*}$} \\
				\bottomrule
			\end{tabular}%
			\begin{tablenotes}
				\item \textbf{Note:} \textbf{Bold} and \underline{underline} denote the best and second-best results. $^{*}$ indicates a highly statistically significant improvement ($p < 0.01$) vs. the second-best method.
			\end{tablenotes}
		\end{threeparttable}%
	}
\end{table*}

\textbf{Generalization to Small Target Domains (ISIC2018 $\to$ PH$^2$):} In this scenario, UGDD-Net achieves an outstanding IoU of 85.93\% and reduces the HD95 to 11.07 pixels. The substantial performance gap between our method and baseline models highlights the capability of our dual-domain architecture to adapt to new clinical environments without catastrophic performance drops.

\textbf{Generalization from Limited Data (PH$^2$ $\to$ ISIC2018):} The superiority of our framework is even more pronounced in this highly challenging setting. Since PH$^2$ is a much smaller dataset, standard models trained on it tend to overfit local spatial biases, leading to severe performance degradation when tested on the highly diverse ISIC2018 dataset. Despite this harsh domain shift, UGDD-Net maintains an impressive Dice score of 83.29\% (a 1.93\% absolute improvement over the second-best UDEL) and reduces the ASSD to 6.64 pixels. Furthermore, our model significantly lowers the Expected Calibration Error (ECE) to 12.02\%. This notably low ECE indicates that UGDD-Net remains reliable and well-calibrated even when facing completely unseen data distributions.

\subsubsection{Qualitative Visualization}
\begin{figure*}[!ht]
	\centering
	\includegraphics[width=0.95\linewidth]{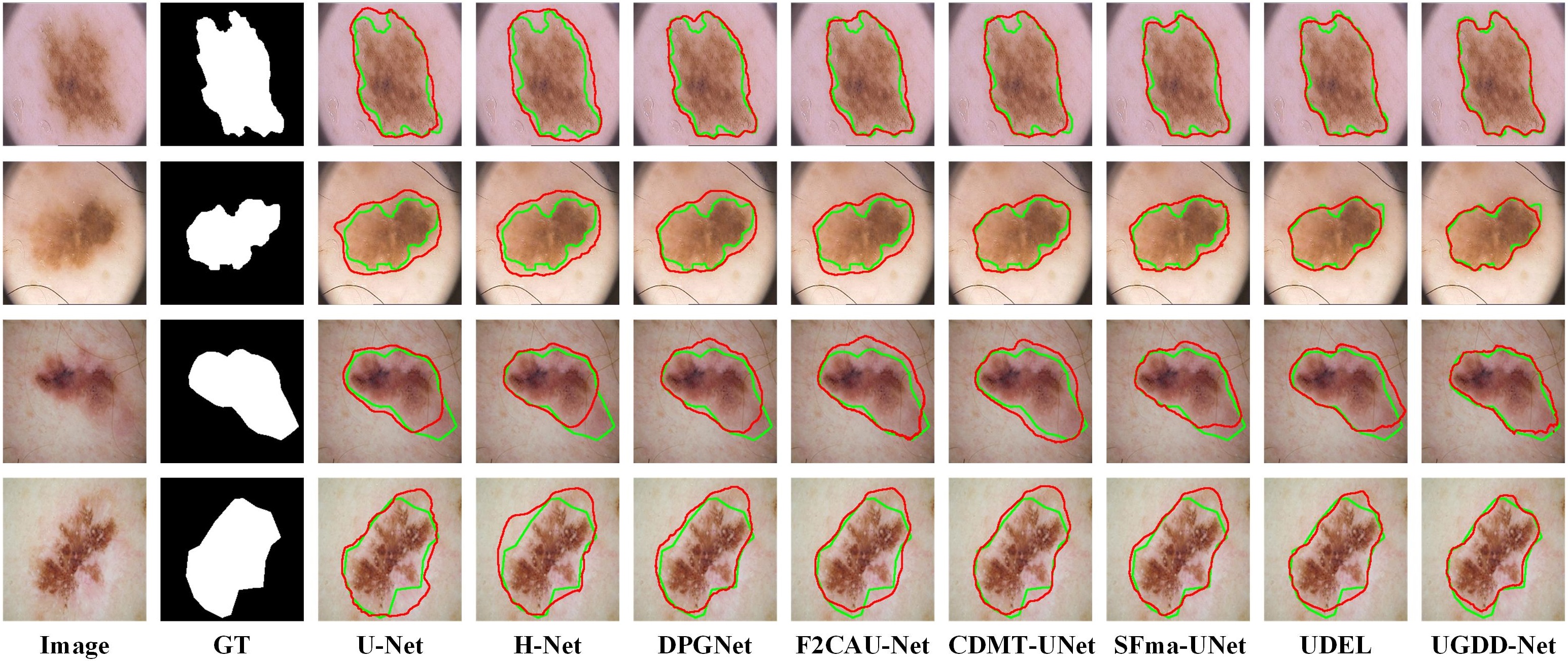}
	\vspace{-1mm} 
	\caption{Qualitative visualization of cross-dataset generalization. The first two rows show the results of models trained on ISIC2018 and tested on PH$^2$ (ISIC2018 $\to$ PH$^2$), while the last two rows show the reverse setting (PH$^2$ $\to$ ISIC2018). The selected challenging samples feature highly irregular shapes and severely blurred boundaries. The green curves outline the ground truth contours, and the red curves indicate the predicted boundaries.}
	\label{fig:cross_visual}
\end{figure*}

To visually demonstrate this robustness, we present the cross-dataset segmentation results in Fig.~\ref{fig:cross_visual}. All showcased examples---whether from ISIC2018 $\to$ PH$^2$ (first two rows) or PH$^2$ $\to$ ISIC2018 (last two rows)---share common clinical difficulties: highly irregular shapes and severely blurred boundaries.

Under cross-domain conditions, the absolute color and contrast distributions of lesions change drastically. Consequently, baseline models that heavily rely on spatial intensity fail to locate exact tissue margins, leading to erratic boundary delineations, manifesting as either severe under-segmentation or over-segmentation. In contrast, UGDD-Net actively resists these domain-specific biases. By extracting domain-invariant structural features through the frequency branch and dynamically calibrating ambiguous edge responses via the UGML module, our model more accurately delineates irregular contours that closely align with the ground truth, highlighting its robust adaptability to domain shifts encountered in clinical practice.

\subsection{Robustness Analysis on Challenging Scenarios}

\subsubsection{Quantitative Analysis}
To comprehensively evaluate the model's performance and its robustness against severe morphological and visual interferences, we conduct experiments on the HAM10000 dataset. In addition to evaluating all samples, we specifically isolate a ``Hard Samples'' subset, which includes cases characterized by small lesions, ambiguous boundaries, irregular topologies, and significant artifacts. The quantitative results are summarized in Table~\ref{tab:ham_fine_grained}.

\begin{table*}[!ht]
	\centering
	\caption{Quantitative comparison of segmentation performance on the HAM10000 dataset. The ``Hard Samples'' subset comprises challenging cases where the baseline U-Net achieves an IoU below its global average ($< 86.75\%$), typically featuring small lesions, ambiguous boundaries, irregular topologies, and significant artifacts. {\sfmark} denotes Spatial-Frequency methods, and {\uncmark} denotes Uncertainty-aware methods.}
	\label{tab:ham_fine_grained}
	
	\newcommand{\spm}{\!\pm\mkern 2mu\scriptstyle} 
	
	\resizebox{\textwidth}{!}{%
		\begin{threeparttable}
			\begin{tabular}{lcccccccccc}
				\toprule
				\multirow{2}{*}{\textbf{Model}} & \multicolumn{5}{c}{\textbf{All Samples}} & \multicolumn{5}{c}{\textbf{Hard Samples}} \\
				\cmidrule(lr){2-6} \cmidrule(lr){7-11}
				& IoU \scriptsize{(\%)} $\uparrow$ & Dice \scriptsize{(\%)} $\uparrow$ & HD95 \scriptsize{(px)} $\downarrow$ & ASSD \scriptsize{(px)} $\downarrow$ & ECE \scriptsize{(\%)} $\downarrow$ & IoU \scriptsize{(\%)} $\uparrow$ & Dice \scriptsize{(\%)} $\uparrow$ & HD95 \scriptsize{(px)} $\downarrow$ & ASSD \scriptsize{(px)} $\downarrow$ & ECE \scriptsize{(\%)} $\downarrow$ \\
				\midrule
				
				U-Net \citep{ronneberger2015u}  & $86.75\spm1.07$ & $92.36\spm0.93$ & $11.42\spm1.33$ & $3.99\spm1.14$ & $12.67\spm1.66$ & $70.71\spm1.27$ & $81.23\spm1.02$ & $25.14\spm1.49$ & $10.56\spm1.30$ & $16.40\spm1.85$ \\
				Swin-Unet \citep{cao2022swin}  & $87.30\spm1.30$ & $92.71\spm1.15$ & $11.81\spm1.65$ & $4.14\spm1.45$ & $16.75\spm2.05$ & $69.77\spm1.55$ & $80.23\spm1.25$ & $25.91\spm1.85$ & $11.23\spm1.65$ & $19.97\spm2.30$ \\
				VM-UNet \citep{ruan2024vm}      & $88.12\spm0.98$ & $93.18\spm0.84$ & $10.96\spm1.21$ & $3.77\spm1.02$ & $13.62\spm1.50$ & $71.99\spm1.15$ & $81.97\spm0.92$ & $23.95\spm1.35$ & $10.11\spm1.16$ & $16.56\spm1.67$ \\
				MPBA-Net \citep{huang2025lesion} & $89.35\spm0.66$ & $93.86\spm0.53$ & $9.97\spm0.77$ & $3.37\spm0.58$ & $14.29\spm0.95$ & $75.82\spm0.75$ & $84.70\spm0.59$ & $21.69\spm0.86$ & $8.87\spm0.66$ & $18.23\spm1.03$ \\
				WinGraphUNet \citep{kui2025wingraphunet}  & $90.44\spm0.61$ & $94.65\spm0.49$ & $8.57\spm0.70$ & $2.98\spm0.52$ & $15.87\spm0.87$ & $76.79\spm0.70$ & $85.52\spm0.55$ & $19.94\spm0.78$ & $7.51\spm0.59$ & $19.44\spm0.94$ \\
				CDMT-UNet\sfmark \citep{cai2026cdmt} & $90.44\spm1.03$ & $94.65\spm0.88$ & $8.93\spm1.27$ & $3.09\spm1.08$ & $15.87\spm1.58$ & $76.79\spm1.21$ & $85.52\spm0.97$ & $20.55\spm1.42$ & $7.87\spm1.23$ & $19.44\spm1.76$ \\
				SkinMamba\sfmark \citep{zou2024skinmamba}   & $88.94\spm0.75$ & $93.64\spm0.62$ & $8.42\spm0.89$ & $2.92\spm0.70$ & $13.51\spm1.11$ & $72.02\spm0.87$ & $82.19\spm0.69$ & $20.04\spm1.00$ & $7.68\spm0.80$ & $16.65\spm1.21$ \\
				DEviS\uncmark \citep{zou2025toward}            & $89.86\spm0.43$ & $94.24\spm0.31$ & $10.48\spm0.45$ & $3.54\spm0.27$ & $11.33\spm0.56$ & $74.75\spm0.47$ & $84.09\spm0.36$ & $22.12\spm0.50$ & $9.24\spm0.31$ & $14.76\spm0.58$ \\
				AHF-U-Net\uncmark \citep{munia2025attention}  & $89.51\spm0.70$ & $94.06\spm0.58$ & $10.23\spm0.83$ & $3.46\spm0.64$ & $12.59\spm1.03$ & $74.26\spm0.81$ & $83.81\spm0.64$ & $22.04\spm0.93$ & $8.99\spm0.73$ & $15.72\spm1.12$ \\
				DPGNet\uncmark \citep{wang2025dpgnet}         & $89.74\spm0.52$ & $94.18\spm0.40$ & $10.03\spm0.58$ & $3.40\spm0.39$ & $12.21\spm0.72$ & $75.14\spm0.58$ & $84.33\spm0.45$ & $21.98\spm0.64$ & $8.94\spm0.45$ & $15.35\spm0.76$ \\
				\midrule
				
				H-Net \citep{zhou2022h} & $87.02\spm0.93$ & $92.45\spm0.80$ & $11.06\spm1.14$ & $3.93\spm0.95$ & $13.02\spm1.42$ & $71.45\spm1.10$ & $81.76\spm0.88$ & $24.47\spm1.28$ & $10.23\spm1.09$ & $16.49\spm1.58$ \\
				BDFormer \citep{ji2025bdformer}  & $89.41\spm1.25$ & $93.93\spm1.11$ & $9.31\spm1.59$ & $3.22\spm1.39$ & $16.09\spm1.97$ & $73.27\spm1.49$ & $82.84\spm1.20$ & $21.35\spm1.78$ & $8.64\spm1.58$ & $19.92\spm2.21$ \\
				CFFormer \citep{li2026cfformer} & $90.27\spm1.21$ & $94.56\spm1.06$ & $10.67\spm1.52$ & $3.67\spm1.33$ & $14.73\spm1.89$ & $76.34\spm1.44$ & $85.27\spm1.16$ & $22.99\spm1.71$ & $9.85\spm1.51$ & $18.80\spm2.12$ \\
				DSU-Net \citep{zhong2025dsu}   & $89.46\spm1.12$ & $93.95\spm0.97$ & $10.83\spm1.40$ & $3.71\spm1.20$ & $14.31\spm1.74$ & $73.72\spm1.32$ & $83.17\spm1.06$ & $23.53\spm1.57$ & $9.98\spm1.37$ & $18.56\spm1.94$ \\
				DMA-Net \citep{zhai2024dma}   & $88.69\spm1.16$ & $93.48\spm1.02$ & $10.47\spm1.46$ & $3.49\spm1.26$ & $13.94\spm1.81$ & $72.58\spm1.38$ & $82.31\spm1.11$ & $22.55\spm1.64$ & $9.64\spm1.44$ & $16.82\spm2.03$ \\
				SF-UNet\sfmark \citep{zhou2024spatial} & $88.03\spm0.89$ & $93.11\spm0.75$ & $9.15\spm1.08$ & $3.16\spm0.89$ & $13.22\spm1.34$ & $73.19\spm1.04$ & $82.72\spm0.83$ & $20.99\spm1.21$ & $8.39\spm1.02$ & $16.28\spm1.49$ \\
				SFma-UNet\sfmark \citep{liu2025sfma}   & $90.83\spm0.80$ & $94.87\spm0.66$ & $8.57\spm0.96$ & $2.98\spm0.77$ & $13.98\spm1.19$ & $76.93\spm0.92$ & $85.66\spm0.73$ & $19.94\spm1.07$ & $7.51\spm0.87$ & $17.15\spm1.30$ \\
				DPMF-Net\sfmark \citep{huang2025dpmf}  & $89.98\spm0.84$ & $94.32\spm0.71$ & $9.04\spm1.02$ & $3.13\spm0.83$ & $13.07\spm1.27$ & $76.66\spm0.98$ & $85.41\spm0.78$ & $20.76\spm1.14$ & $8.24\spm0.94$ & $16.37\spm1.39$ \\
				F2CAU-Net\uncmark \citep{zhou2025f2cau}  & $90.05\spm0.57$ & $94.36\spm0.44$ & $9.32\spm0.64$ & $3.18\spm0.46$ & $11.67\spm0.80$ & $76.15\spm0.64$ & $84.99\spm0.50$ & $21.03\spm0.71$ & $8.51\spm0.52$ & $14.88\spm0.85$ \\
				UDEL\uncmark \citep{wang2025udel}   & \underline{$91.07\spm0.48$} & \underline{$94.98\spm0.35$} & \underline{$8.16\spm0.51$} & \underline{$2.88\spm0.33$} & \underline{$11.02\spm0.64$} & \underline{$77.10\spm0.53$} & \underline{$85.79\spm0.41$} & \underline{$18.96\spm0.57$} & \underline{$7.31\spm0.38$} & \underline{$14.35\spm0.67$} \\
				\midrule
				
				\textbf{UGDD-Net (Ours)}\bothmark & $\mathbf{92.61}\spm\mathbf{0.40}$\rlap{$^{*}$} & $\mathbf{96.01}\spm\mathbf{0.27}$\rlap{$^{*}$} & $\mathbf{6.45}\spm\mathbf{0.41}$\rlap{$^{*}$} & $\mathbf{2.07}\spm\mathbf{0.22}$\rlap{$^{*}$} & $\mathbf{8.53}\spm\mathbf{0.49}$\rlap{$^{*}$} & $\mathbf{79.92}\spm\mathbf{0.44}$\rlap{$^{*}$} & $\mathbf{87.78}\spm\mathbf{0.31}$\rlap{$^{*}$} & $\mathbf{17.03}\spm\mathbf{0.46}$\rlap{$^{*}$} & $\mathbf{6.28}\spm\mathbf{0.25}$\rlap{$^{*}$} & $\mathbf{11.72}\spm\mathbf{0.51}$\rlap{$^{*}$} \\
				\bottomrule
			\end{tabular}%
			
			\begin{tablenotes}
				\item \textbf{Note:} \textbf{Bold} and \underline{underline} denote the best and second-best results. $^{*}$ indicates a highly statistically significant improvement ($p < 0.01$) vs. the second-best method.
			\end{tablenotes}
			
		\end{threeparttable}%
	}
\end{table*}

\textbf{Performance on All Samples:} On the complete dataset, UGDD-Net establishes a new state-of-the-art. It achieves an impressive IoU of 92.61\% and a Dice score of 96.01\%, outperforming the second-best method (UDEL) while achieving a remarkably low ECE of 8.53\%. This indicates that our model provides highly accurate and well-calibrated predictions under standard conditions.

\textbf{Robustness on Hard Samples:} The true advantage of UGDD-Net emerges in the challenging subset. As shown in Table~\ref{tab:ham_fine_grained}, baseline models and even advanced spatial-frequency or uncertainty-aware methods experience a sharp performance drop when facing hard samples. For instance, the state-of-the-art uncertainty method UDEL's IoU drops to 77.10\% and its HD95 increases to 18.96 pixels. In comparison, UGDD-Net maintains a significantly higher IoU of 79.92\% and limits the HD95 to 17.03 pixels. Furthermore, our method reduces the ASSD to 6.28 pixels (a 1.03-pixel improvement over UDEL) and keeps the ECE at 11.72\%. This direct comparison demonstrates that the uncertainty-guided dual-domain fusion mechanism effectively prevents the model from collapsing under severe interference.

\subsubsection{Qualitative Visualization}
\begin{figure*}[!ht]
	\centering
	\includegraphics[width=0.95\linewidth]{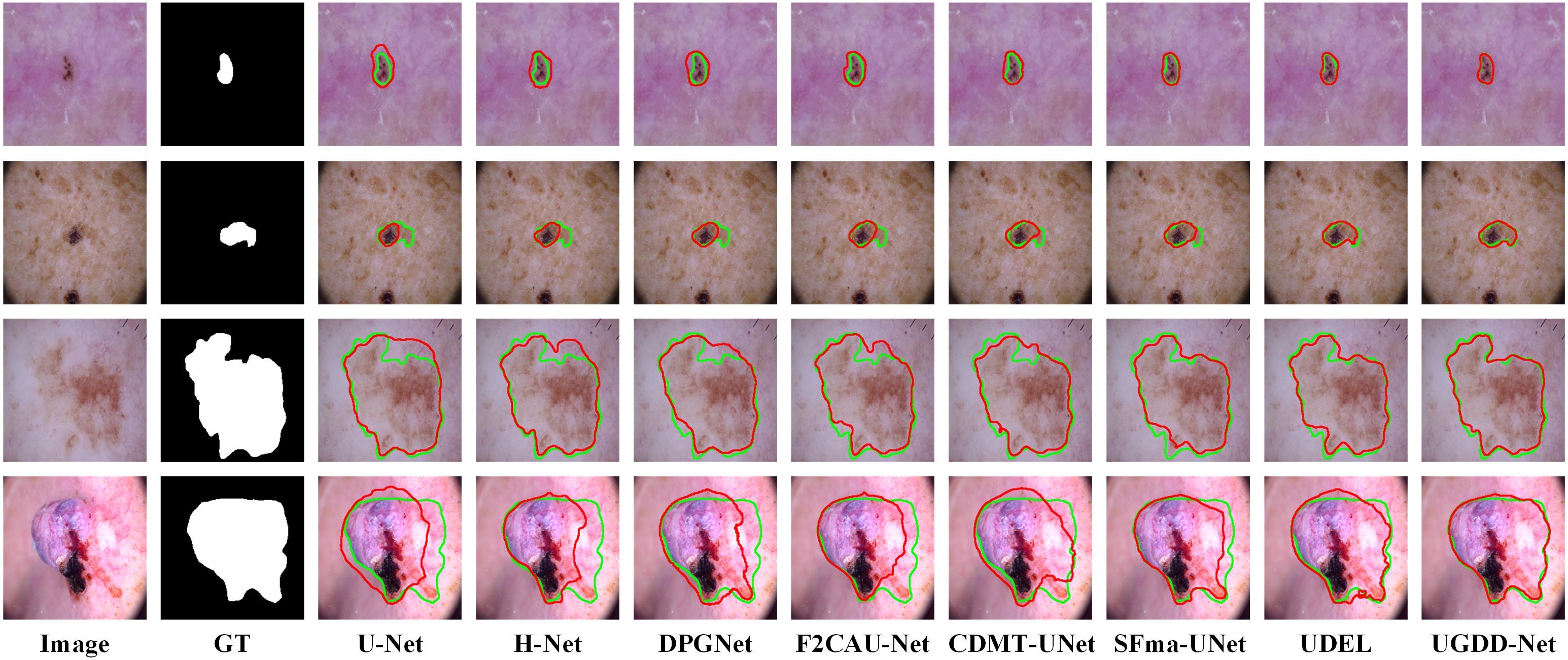}
	\caption{Qualitative visualization of challenging scenarios on the HAM10000 dataset. The samples represent four typical difficulties: small lesions with blurry edges (1st row), small irregular targets (2nd row), irregular shapes with blurry boundaries (3rd row), and lesions with reflective artifacts (4th row). The green curves outline the ground truth contours, and the red curves indicate the predicted boundaries.}
	\label{fig:ham_visual}
\end{figure*}

To intuitively illustrate the performance on these challenging cases, Fig.~\ref{fig:ham_visual} visualizes four typical hard samples from the HAM10000 dataset.

The first two rows display small targets accompanied by blurry edges (first row) and irregular morphological structures (second row). Existing comparative methods tend to lose the precise spatial details of such small targets during deep feature extraction, often resulting in inaccurate segmentation with noticeable under-segmentation or over-segmentation errors. UGDD-Net, leveraging precise high-frequency contour perception, accurately anchors these micro-lesions and delineates their exact boundaries.

The last two rows exhibit lesions with highly irregular shapes, further complicated by severe boundary blurriness (third row) and strong reflective artifacts (fourth row). In these scenarios, comparative networks struggle to distinguish true lesion tissue from artifacts, leading to distorted contours or erroneous expansion into the surrounding skin. Driven by the UGBFF module, our model actively estimates the high uncertainty of these reflective artifacts and dynamically filters out the interference, yielding highly accurate segmentation masks.

\subsection{Ablation study for UGDD-Net}
\label{subsec:ablation}

\begin{figure*}[!ht]
	\centering
	\includegraphics[width=0.95\textwidth]{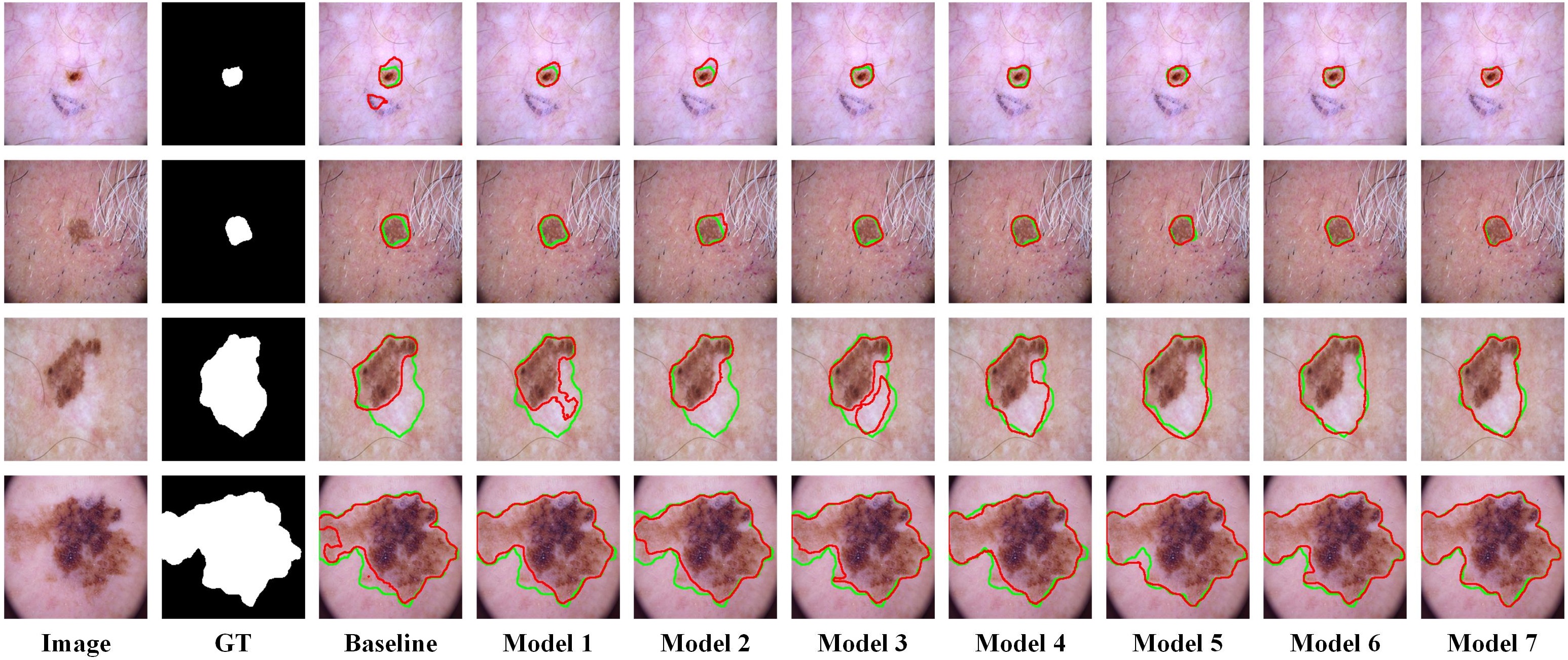}
	\caption{Qualitative visualization of the ablation study on the HAM10000 dataset. The rows display challenging samples: small targets disturbed by artifacts such as blue markers and dense hair (1st and 2nd rows), and highly irregular lesions with ambiguous boundaries (3rd and 4th rows). The columns illustrate the progressive evolution of segmentation quality from the Baseline (single-domain without uncertainty guidance) to the full UGDD-Net (Model 7). The green curves outline the ground truth contours, and the red curves indicate the predicted boundaries.}
	\label{fig:ablation_vis}
\end{figure*}

To rigorously validate the contribution of each proposed component in UGDD-Net, we conduct a comprehensive ablation study on the HAM10000 dataset. We progressively construct our network from a standard U-Net baseline and evaluate the performance on both the complete dataset (All Samples) and the challenging subset (Hard Samples). The quantitative results are detailed in Table~\ref{tab:ablation}, and the corresponding qualitative visualizations are presented in Fig.~\ref{fig:ablation_vis}.

\begin{table*}[!ht]
	\centering
	\caption{Quantitative ablation study of UGDD-Net components on the HAM10000 dataset. Baseline refers to the standard U-Net. DD: Dual-Domain; UGBFF: Uncertainty-Guided Bi-directional Feature Fusion; UGGR: Uncertainty-Guided Graph Refinement; UGML: Uncertainty-Guided Margin-Adaptive Loss. Model 7 represents our full proposed UGDD-Net.}
	\label{tab:ablation}
	\newcommand{\spm}{\!\pm\!} 
	
	\resizebox{\textwidth}{!}{%
		\begin{threeparttable}
			\begin{tabular}{l c ccc ccccc ccccc}
				\toprule
				\multirow{2}{*}{\textbf{Model}} & \multirow{2}{*}{\textbf{DD Mode}} & \multicolumn{3}{c}{\textbf{Components}} & \multicolumn{5}{c}{\textbf{All Samples}} & \multicolumn{5}{c}{\textbf{Hard Samples}} \\
				\cmidrule(lr){3-5} \cmidrule(lr){6-10} \cmidrule(lr){11-15}
				& & {UGBFF} & {UGGR} & {UGML} & {IoU\,\scriptsize{(\%)}\,$\uparrow$} & {Dice\,\scriptsize{(\%)}\,$\uparrow$} & {HD95\,\scriptsize{(px)}\,$\downarrow$} & {ASSD\,\scriptsize{(px)}\,$\downarrow$} & {ECE\,\scriptsize{(\%)}\,$\downarrow$} & {IoU\,\scriptsize{(\%)}\,$\uparrow$} & {Dice\,\scriptsize{(\%)}\,$\uparrow$} & {HD95\,\scriptsize{(px)}\,$\downarrow$} & {ASSD\,\scriptsize{(px)}\,$\downarrow$} & {ECE\,\scriptsize{(\%)}\,$\downarrow$} \\
				\midrule
				
				\textbf{Baseline} & Single (S) & -- & -- & -- & 86.75 & 92.36 & 11.42 & 3.99 & 12.67 & 70.71 & 81.23 & 25.14 & 10.56 & 16.40 \\
				\midrule 
				
				Model 1 & S \& S & -- & -- & -- & 88.36 & 93.34 & 10.06 & 3.47 & 11.96 & 72.64 & 82.57 & 22.55 & 8.89 & 15.58 \\
				Model 2 & F \& F & -- & -- & -- & 88.01 & 93.15 & 10.38 & 3.67 & 12.32 & 72.16 & 82.25 & 23.18 & 9.26 & 16.01 \\
				Model 3 & S \& F & -- & -- & -- & 89.92 & 94.34 & 8.74 & 2.96 & 11.93 & 74.70 & 84.03 & 20.92 & 8.23 & 15.45 \\
				\midrule 
				
				Model 4 & S \& F & \checkmark & -- & -- & 90.73 & 94.82 & 8.04 & 2.71 & 10.98 & 76.65 & 85.41 & 19.26 & 7.37 & 14.42 \\
				Model 5 & S \& F & -- & \checkmark & -- & 90.91 & 94.95 & 7.89 & 2.65 & 10.84 & 76.89 & 85.58 & 19.07 & 7.29 & 14.26 \\
				Model 6 & S \& F & \checkmark & \checkmark & -- & 91.55 & 95.35 & 7.36 & 2.43 & 10.39 & 78.35 & 86.64 & 18.11 & 6.82 & 13.81 \\
				\midrule
				
				\textbf{Model 7} & \textbf{S \& F} & \checkmark & \checkmark & \checkmark & \textbf{92.61} & \textbf{96.01} & \textbf{6.45} & \textbf{2.07} & \textbf{8.53} & \textbf{79.92} & \textbf{87.78} & \textbf{17.03} & \textbf{6.28} & \textbf{11.72} \\
				
				\bottomrule
			\end{tabular}
		\end{threeparttable}%
	}
\end{table*}

\textbf{Effectiveness of the Dual-Domain Architecture:} 
We first evaluate the necessity of combining spatial and frequency domains. As shown in Table~\ref{tab:ablation}, expanding the single-path Baseline into a naive dual-spatial (Model 1, S\&S) or dual-frequency (Model 2, F\&F) architecture yields only marginal performance gains. However, when coupling the spatial and frequency domains (Model 3, S\&F), the IoU on Hard Samples increases significantly to 74.70\% (a 3.99\% absolute improvement over the Baseline). This validates our core motivation: spatial semantics and high-frequency details contain highly complementary information that a homogeneous dual-path network cannot capture.

\textbf{Effectiveness of Uncertainty-Guided Components:} 
Building upon the S\&F architecture (Model 3), we further analyze the impact of our uncertainty-guided modules. Unlike traditional paradigms, these modules transform uncertainty from a passive metric into an active guiding signal, enabling the model to self-diagnose and self-rectify during the learning process:
\begin{itemize}
	\item \textbf{UGBFF and UGGR:} Integrating the UGBFF module (Model 4) boosts the Hard Sample IoU to 76.65\%. This confirms that UGBFF effectively resolves the ``blind fusion'' dilemma. By utilizing pixel-level uncertainty to dynamically modulate cross-domain interactions, it ensures that high-confidence representations actively supervise unreliable features. Independently adding the UGGR module (Model 5) yields an IoU of 76.89\%, validating its ability to overcome ``passive reasoning.'' By constructing a sparse graph for topology-aware relational inference, UGGR actively propagates reliable semantic consensus to rectify uncertain query nodes. When both modules are combined (Model 6), they complement each other effectively, raising the Hard Sample IoU to 78.35\% and reducing the HD95 to 18.11 pixels.
	
	\item \textbf{UGML:} Finally, applying the Uncertainty-Guided Margin-Adaptive Loss (Model 7, our full UGDD-Net) overcomes the limitation of ``rigid optimization.'' This objective dynamically adapts decision margins based on pixel-wise uncertainty---enforcing strict constraints on confident pixels while relaxing penalties on uncertain ones. Consequently, this final addition not only pushes the Hard Sample IoU to an impressive 79.92\%, but also drastically reduces the Expected Calibration Error (ECE) to 11.72\%. This demonstrates that UGML prevents the over-penalization of inherently ambiguous regions, effectively suppressing erratic confidence fluctuations to achieve superior statistical calibration.
\end{itemize}

\textbf{Qualitative Visualization:} 
Fig.~\ref{fig:ablation_vis} intuitively corroborates this progressive improvement. The first two rows display small targets heavily disturbed by artifacts (e.g., blue markers and dense hair), while the last two rows present highly irregular targets with blurry boundaries. The Baseline and early naive dual-domain networks (Models 1--3) struggle to accurately delineate these difficult targets, resulting in over-segmentation or under-segmentation that fails to capture irregular lesion extensions. With the progressive integration of uncertainty guidance (Models 4--7), the network gradually learns to strictly modulate cross-domain feature fusion based on local reliability and propagate high-confidence semantic labels to rectify ambiguous predictions. Ultimately, the full UGDD-Net (Model 7) yields highly accurate and tight boundaries that align closely with the ground truth. Notably, the visual superiority of our method is most prominent in cases of severe boundary ambiguity and irregular topologies (last two rows), where UGDD-Net successfully recovers complex morphological structures that the baseline and intermediate variants largely miss.

\subsection{Generalizability Across Different Backbones}
\label{subsec:backbones}

A hallmark of a robust architectural paradigm is its generalizability across different network feature extractors. To demonstrate that our uncertainty-guided dual-domain framework is highly adaptable, we evaluate its compatibility with diverse modern architectures, including CNN (U-Net), Transformer (Swin-Unet), and Mamba (VM-UNet). We construct five variants (Models I--V) by alternating the feature extractors in the spatial and frequency streams while keeping all uncertainty-guided modules active. The 5-fold cross-validation results are presented in Table~\ref{tab:backbones}.

\begin{table*}[!ht]
	\centering
	\caption{Segmentation performance of different backbone combinations on the HAM10000 dataset. All models are equipped with the proposed uncertainty modules. The heterogeneous CNN-Mamba architecture (Model IV) achieves the best performance.}
	\label{tab:backbones}
	\newcommand{\spm}{\!\pm\!} 
	
	\resizebox{\textwidth}{!}{%
		\begin{tabular}{l ll ccccc ccccc}
			\toprule
			
			\multirow{2}{*}{\textbf{Model}} & \multicolumn{2}{c}{\textbf{Stream Configuration}} & \multicolumn{5}{c}{\textbf{All Samples}} & \multicolumn{5}{c}{\textbf{Hard Samples}} \\
			
			\cmidrule(lr){2-3} \cmidrule(lr){4-8} \cmidrule(lr){9-13}
			
			& {Spatial} & {Frequency} & {IoU\,\scriptsize{(\%)}\,$\uparrow$} & {Dice\,\scriptsize{(\%)}\,$\uparrow$} & {HD95\,\scriptsize{(px)}\,$\downarrow$} & {ASSD\,\scriptsize{(px)}\,$\downarrow$} & {ECE\,\scriptsize{(\%)}\,$\downarrow$} & {IoU\,\scriptsize{(\%)}\,$\uparrow$} & {Dice\,\scriptsize{(\%)}\,$\uparrow$} & {HD95\,\scriptsize{(px)}\,$\downarrow$} & {ASSD\,\scriptsize{(px)}\,$\downarrow$} & {ECE\,\scriptsize{(\%)}\,$\downarrow$} \\
			\midrule
			
			Model I & U-Net & U-Net & 92.61 & 96.01 & 6.45 & 2.07 & 8.53 & 79.92 & 87.78 & 17.03 & 6.28 & 11.72 \\
			Model II & Swin-Unet & Swin-Unet & 91.99 & 95.54 & 7.15 & 2.39 & 9.21 & 78.81 & 86.87 & 17.79 & 6.71 & 12.43 \\
			Model III & U-Net & Swin-Unet & 92.88 & 96.21 & 6.16 & 1.93 & 8.56 & 80.41 & 88.19 & 16.68 & 6.09 & 11.78 \\
			
			\textbf{Model IV} & \textbf{U-Net} & \textbf{VM-UNet} & \textbf{93.19} & \textbf{96.39} & \textbf{5.83} & \textbf{1.82} & \textbf{8.49} & \textbf{80.97} & \textbf{88.62} & \textbf{16.32} & \textbf{5.90} & \textbf{11.67} \\
			
			Model V & Swin-Unet & VM-UNet & 92.29 & 95.80 & 6.82 & 2.23 & 8.94 & 79.33 & 87.31 & 17.45 & 6.52 & 12.12 \\
			
			\bottomrule
		\end{tabular}%
	}
\end{table*}

\textbf{Homogeneous Configurations:} We first compare models using identical architectures for both streams. Model I (CNN + CNN) serves as our default UGDD-Net and demonstrates strong performance. Interestingly, replacing the CNNs with Transformers (Model II, Swin-Unet + Swin-Unet) results in a performance decrease, with the Hard Sample IoU shifting from 79.92\% to 78.81\%. This indicates that for skin lesion segmentation, where fuzzy boundaries demand precise pixel-level localization, the local inductive biases of CNNs prove advantageous over pure self-attention mechanisms.

\textbf{Heterogeneous Configurations:} The extensibility of our framework is further revealed when mixing different architectural paradigms. When coupling a CNN-based spatial stream with a modern global extractor in the frequency stream (Model III and Model IV), the performance consistently surpasses the homogeneous baselines. Most notably, Model IV (U-Net Spatial + VM-UNet Frequency) achieves the highest overall performance, pushing the Hard Sample IoU to 80.97\% and reducing the HD95 to 16.32 pixels. 

This superior result demonstrates that our dual-domain paradigm integrates well with heterogeneous architectures. While the CNN effectively captures local boundary textures in the spatial stream, leveraging Mamba's linear-complexity global modeling in the frequency stream further enhances the extraction of long-range structural dependencies. This analysis proves that our framework is highly flexible and that heterogeneous combinations (e.g., CNN + Mamba) can fully unlock the potential of dual-domain feature representations.

\textbf{Rationale for the Default Configuration:} 
Despite the superior performance of the heterogeneous configuration (Model IV), we establish the homogeneous U-Net architecture (Model I) as the default UGDD-Net throughout this paper. The primary rationale is to maintain architectural simplicity and ensure a fair baseline comparison. By employing a standard, widely adopted backbone, we unambiguously isolate the performance gains attributed to our proposed uncertainty-guided dual-domain mechanisms, proving that the improvements stem directly from our architectural innovations rather than the heavy feature-extraction capabilities of modern backbones like Mamba. Furthermore, this design ensures that UGDD-Net remains computationally efficient and broadly accessible without losing its core effectiveness.

\subsection{Effectiveness of the Proposed Loss Function}
\label{subsec:loss_function}

\begin{figure*}[!ht]
	\centering
	\includegraphics[width=0.95\linewidth]{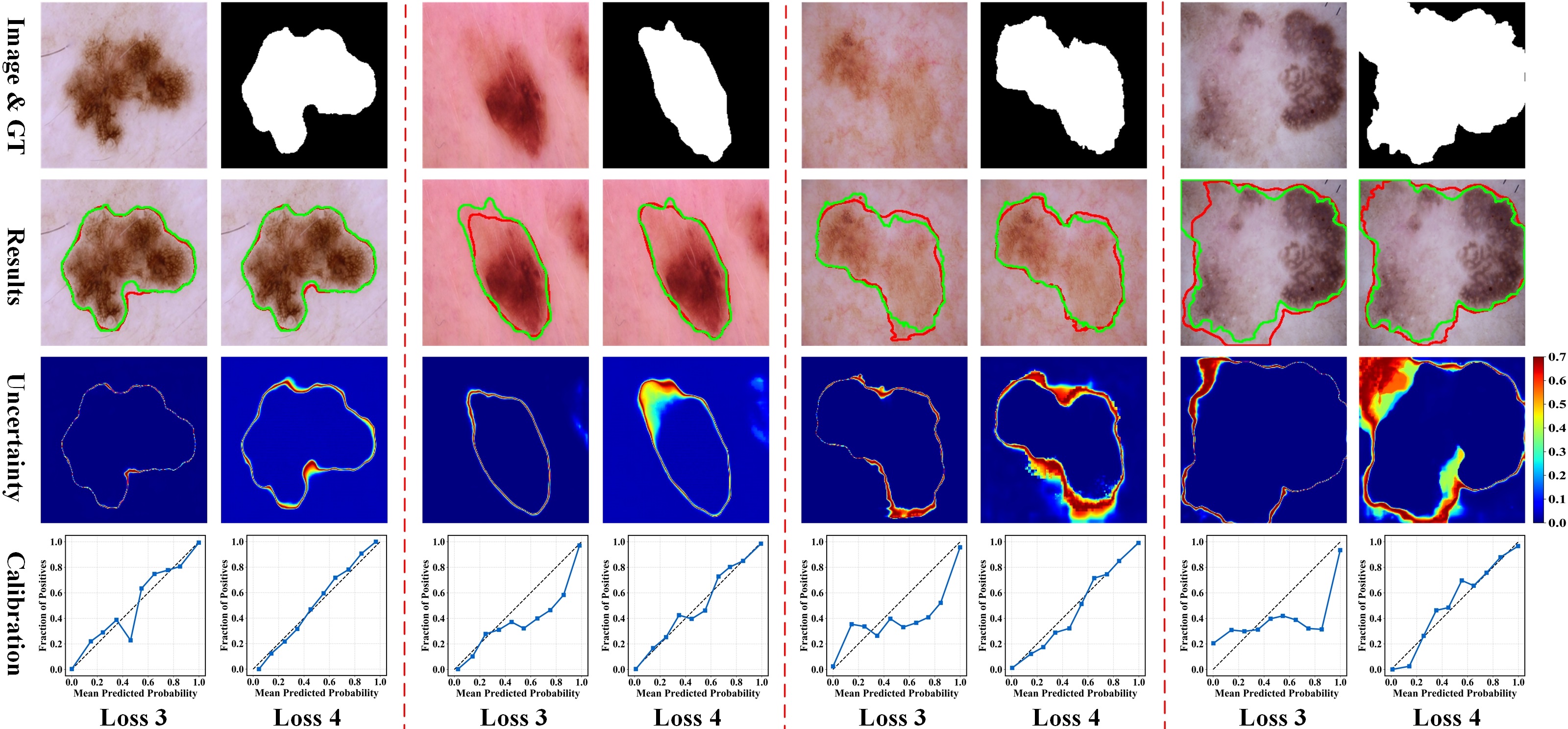}
	\caption{Qualitative and calibration comparison between the Dice + CE + Boundary loss (Loss 3) and our proposed UGML (Loss 4). The figure displays four challenging samples separated by dashed red lines. Each sample is divided into two columns comparing Loss 3 (left) and UGML (right). [Row 1 (Image \& GT)]: Input image and ground truth (GT) mask. [Row 2 (Results)]: Predicted contours (red) overlaid against the GT (green). [Row 3 (Uncertainty)]: Estimated uncertainty maps. [Row 4 (Calibration)]: Corresponding reliability diagrams, where the dashed diagonal line represents perfect calibration. Overall, UGML consistently yields more precise boundaries, diagnostically meaningful uncertainty that precisely highlights intrinsically blurred regions, and superior calibration curves compared to the rigid baseline.}
	\label{fig:calibration}
\end{figure*}

To validate the superiority of the Uncertainty-Guided Margin-Adaptive Loss (UGML), we compare it against widely used segmentation objectives: Dice loss (Loss 1), a combination of Dice and Cross-Entropy (Loss 2), and a boundary-aware variant (Loss 3: Dice + CE + Boundary) \citep{kervadec2019boundary}. To ensure a fair comparison, all other network configurations and uncertainty-guided modules remain identical to our default UGDD-Net; only the loss function is varied. The quantitative results are detailed in Table~\ref{tab:loss_comparison}, and the qualitative visual analysis is presented in Fig.~\ref{fig:calibration}.

\begin{table*}[!ht]
	\centering
	\caption{Segmentation performance and calibration comparison of different loss functions on the HAM10000 dataset. Loss 3 incorporates a boundary-aware penalty \citep{kervadec2019boundary}, while Loss 4 is our proposed UGML.}
	\label{tab:loss_comparison}
	\newcommand{\spm}{\!\pm\!} 
	
	\resizebox{\textwidth}{!}{%
		\begin{tabular}{l ccccc ccccc}
			\toprule
			\multirow{2}{*}{\textbf{Loss Function}} & \multicolumn{5}{c}{\textbf{All Samples}} & \multicolumn{5}{c}{\textbf{Hard Samples}} \\
			\cmidrule(lr){2-6} \cmidrule(lr){7-11}
			& {IoU\,\scriptsize{(\%)}\,$\uparrow$} & {Dice\,\scriptsize{(\%)}\,$\uparrow$} & {HD95\,\scriptsize{(px)}\,$\downarrow$} & {ASSD\,\scriptsize{(px)}\,$\downarrow$} & {ECE\,\scriptsize{(\%)}\,$\downarrow$} & {IoU\,\scriptsize{(\%)}\,$\uparrow$} & {Dice\,\scriptsize{(\%)}\,$\uparrow$} & {HD95\,\scriptsize{(px)}\,$\downarrow$} & {ASSD\,\scriptsize{(px)}\,$\downarrow$} & {ECE\,\scriptsize{(\%)}\,$\downarrow$} \\
			\midrule
			
			\textbf{Loss 1}: Dice & 91.10 & 95.09 & 7.49 & 2.48 & 10.61 & 77.68 & 86.16 & 18.25 & 6.91 & 14.09 \\
			\textbf{Loss 2}: Dice + CE & 91.55 & 95.35 & 7.36 & 2.43 & 10.39 & 78.35 & 86.64 & 18.11 & 6.82 & 13.81 \\
			\textbf{Loss 3}: Dice + CE + Boundary & 91.84 & 95.52 & 7.05 & 2.30 & 10.14 & 78.78 & 86.95 & 17.77 & 6.65 & 13.45 \\
			\textbf{Loss 4: UGML (Ours)} & \textbf{92.61} & \textbf{96.01} & \textbf{6.45} & \textbf{2.07} & \textbf{8.53} & \textbf{79.92} & \textbf{87.78} & \textbf{17.03} & \textbf{6.28} & \textbf{11.72} \\
			
			\bottomrule
		\end{tabular}%
	}
\end{table*}

\textbf{Quantitative Analysis:} As demonstrated in Table~\ref{tab:loss_comparison}, while integrating a boundary-aware penalty (Loss 3) brings incremental improvements over the Dice + CE combination (Loss 2), it still suffers from the limitation of ``rigid optimization.'' Conventional objectives impose uncompromising deterministic penalties based on static labels, which inadvertently forces the network to overfit subjective label noise. In contrast, our proposed UGML (Loss 4) achieves notable gains across all metrics. On the challenging Hard Samples subset, UGML pushes the IoU to 79.92\% and reduces the HD95 to 17.03 pixels. Crucially, UGML drastically reduces the Expected Calibration Error (ECE) to 11.72\%. This consistent improvement confirms that UGML effectively calibrates the model to be trustworthy, mitigating the severe miscalibration often caused by rigid objective functions.

\textbf{Qualitative and Interpretability Analysis:} The visual comparison in Fig.~\ref{fig:calibration} provides compelling evidence for how UGML overcomes rigid optimization by dynamically adapting decision margins based on pixel-wise uncertainty. We compare the segmentation results, uncertainty maps, and calibration curves between Loss 3 and our UGML across four irregular lesion samples:
\begin{itemize}
	\item \textbf{Segmentation Accuracy (Row 2):} Loss 3 aggressively penalizes inherently ambiguous pixels, which forces the network to make hard decisions in uncertain regions. This rigid supervision leads to erratic boundary delineations, manifesting as over-segmentation (in Samples 1 and 3) and under-segmentation (in Samples 2 and 4). UGML, by relaxing penalties on uncertain pixels, prevents overfitting to diffuse boundaries and yields tighter, more accurate contours (red lines) that align precisely with the ground truth (green lines).
	
	\item \textbf{Uncertainty Estimation (Row 3):} The estimated uncertainty maps reveal the core mechanism of UGML. For Loss 3, the model yields overly restricted, narrow contour responses. This indicates that the rigid optimization objective severely bottlenecks the model's capacity to express broad ambiguity, seemingly just reacting to localized segmentation errors rather than identifying the true diagnostic difficulty of diffuse regions. Conversely, UGML enforces strict constraints on confident lesion interiors and background regions (showing pure blue), while its uncertainty precisely highlights the intrinsically blurred regions of the lesion (showing broad and distinct heat bands). This accurately mirrors the visual ambiguity in the raw images, providing diagnostically meaningful insights to guide clinical attention.
	
	\item \textbf{Reliability Diagrams (Row 4):} This diagnostically meaningful uncertainty naturally translates to superior statistical calibration. Unlike Loss 3, which produces erratic curves that fluctuate wildly across the diagonal (indicating alternating under- and over-confidence), the reliability curves of UGML tightly hug the ideal reference line. Because the rigid supervision of Loss 3 stifles the reliable representation of ambiguity, its forced pixel-wise predictions are frequently incorrect yet stubbornly confident. UGML effectively mitigates these erratic statistical deviations, delivering consistently reliable and well-calibrated predictions. This indicates that the model's predicted confidence provides a highly trustworthy reference for assessing diagnostic risks.
\end{itemize}

\subsection{Clinical Interpretability of Uncertainty Estimates}
\label{subsec:clinical}

\begin{figure*}[!ht]
	\centering
	\includegraphics[width=0.95\linewidth]{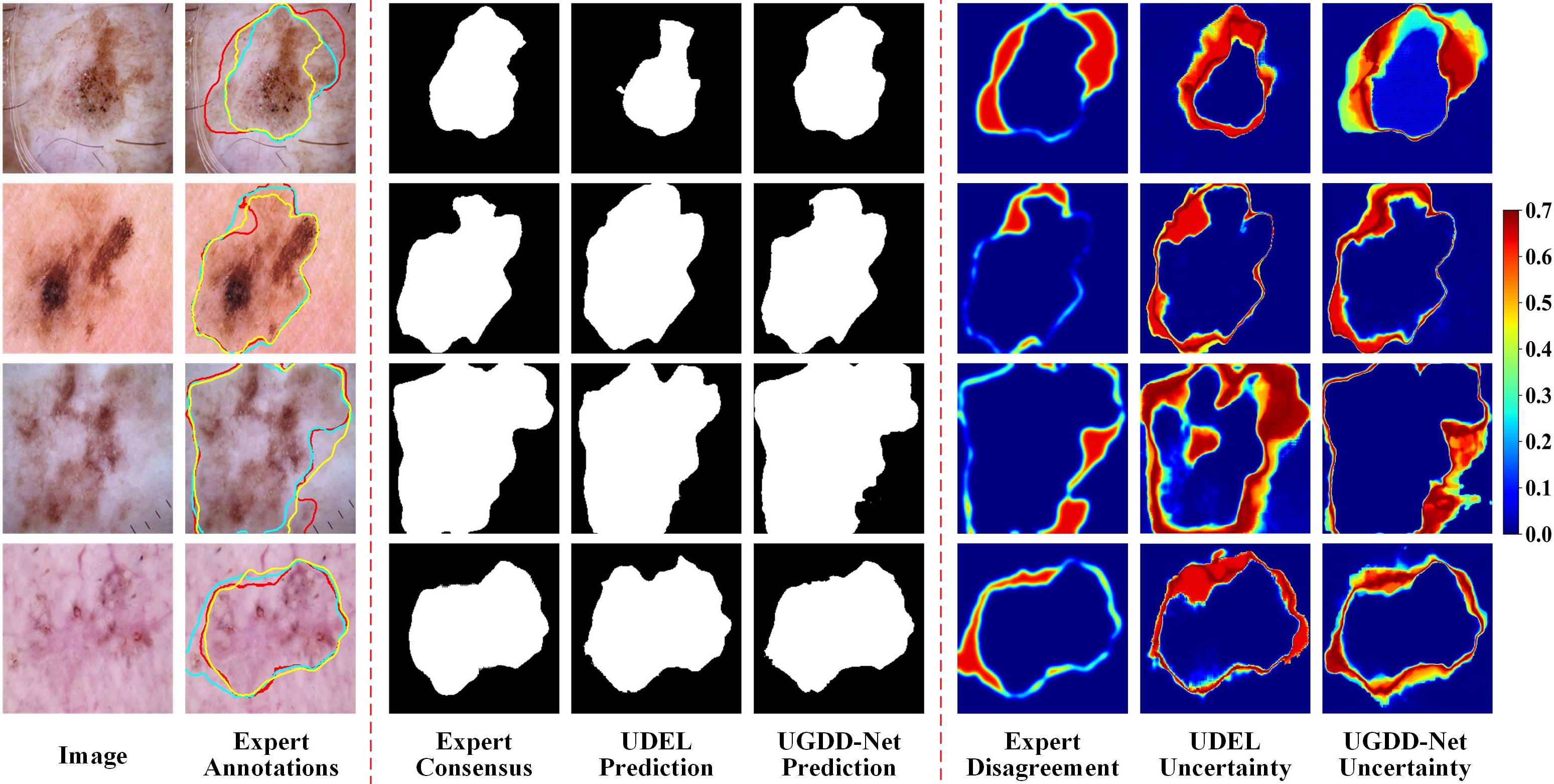}
	\caption{Visual comparison of clinical expert disagreement versus model-predicted uncertainty across four ambiguous lesions. [Cols 1-2]: Original image and three independent expert annotations (shown in different colors). [Cols 3-5]: Expert Consensus mask, alongside the segmentation predictions of UDEL and our proposed UGDD-Net. [Cols 6-8]: Expert Disagreement Map (derived from annotation variance, where red indicates high variability), followed by the estimated uncertainty maps from UDEL and UGDD-Net. Unlike UDEL, the uncertainty patterns predicted by UGDD-Net largely correspond to the actual clinical disagreement, effectively capturing diagnostically ambiguous regions and demonstrating reliable clinical interpretability.}
	\label{fig:expert_alignment}
\end{figure*}

In medical image computing, uncertainty should not merely be a post-hoc mathematical artifact, but rather a diagnostically meaningful indicator of clinical ambiguity. To establish a rigorous ground truth for evaluating this, we invited three board-certified dermatologists to re-annotate 100 highly ambiguous and irregular images from the HAM10000 dataset. By evaluating UGDD-Net's alignment with this actual clinical inter-observer variability, we demonstrate that our framework yields uncertainty maps that possess high clinical relevance. The quantitative evaluation is presented in Table~\ref{tab:expert_alignment}, and the visual comparison is shown in Fig.~\ref{fig:expert_alignment}.

\begin{table}[!ht]
	\centering
	\caption{Quantitative evaluation of segmentation performance and uncertainty alignment. We evaluate segmentation accuracy against the Expert Consensus, and assess the consistency between the estimated uncertainty maps and the Inter-observer Variability derived from three expert annotations.}
	\label{tab:expert_alignment}
	\begin{tabular}{l ccccc}
		\toprule
		& \multicolumn{2}{c}{\textbf{vs. Expert Consensus}} & \multicolumn{3}{c}{\textbf{vs. Inter-observer Variability}} \\ \cmidrule(lr){2-3} \cmidrule(lr){4-6}
		\textbf{Method} & \textbf{IoU}$_{con}$ (\%) $\uparrow$ & \textbf{Dice}$_{con}$ (\%) $\uparrow$ & \textbf{Kappa} $\uparrow$ & \textbf{Pearson} $\uparrow$ & \textbf{Dice}$_{unc}$ (\%) $\uparrow$ \\
		\midrule
		U-Net + MCD & 63.13 & 74.98 & 0.6003 & 0.4345 & 61.44 \\
		UDEL + TTA & 70.21 & 80.23 & 0.6534 & 0.4876 & 67.29 \\
		\textbf{UGDD-Net} & \textbf{73.91} & \textbf{83.37} & \textbf{0.6906} & \textbf{0.5291} & \textbf{71.56} \\
		\bottomrule
	\end{tabular}
\end{table}

\textbf{Quantitative Alignment with Experts:} We benchmarked our model against conventional uncertainty estimation methods (U-Net + MCD) and recent evidential learning approaches (UDEL + TTA). As detailed in Table~\ref{tab:expert_alignment}, UGDD-Net not only achieves the highest segmentation accuracy against the Expert Consensus ($\mathrm{IoU}_{con}$ of 73.91\% and $\mathrm{Dice}_{con}$ of 83.37\%), but more importantly, it exhibits the strongest correlation with human expert disagreement. By achieving a Cohen's Kappa of 0.6906, a Pearson correlation of 0.5291, and a $\mathrm{Dice}_{unc}$ of 71.56\%, UGDD-Net demonstrates that its internal uncertainty explicitly captures the regions where dermatologists genuinely disagree.

\textbf{Visual and Clinical Interpretation:} Fig.~\ref{fig:expert_alignment} provides a comprehensive visual breakdown of this phenomenon across four highly ambiguous and irregular images. We compare the raw images, expert annotations, expert consensus, and expert disagreement maps against the predictions and uncertainty maps of UDEL and our UGDD-Net.
\begin{itemize}
	\item \textbf{Expert Variability (Cols 1-2 \textnormal{\&} 6)}: The diverse annotations from the three experts (Col 2) highlight the inherent subjectivity in lesion boundary delineation. The Expert Disagreement maps (Col 6) reveal that clinical ambiguity is largely concentrated along the highly irregular, fuzzy boundaries.
	
	\item \textbf{Robust Segmentation (Cols 3-5)}: In these highly subjective regions, comparative methods like UDEL (Col 4) often struggle, producing erratic boundaries that deviate from the Expert Consensus (Col 3). In contrast, UGDD-Net (Col 5) delivers more consistent predictions. This visual consistency validates the effectiveness of our core modules: UGBFF employs pixel-level uncertainty to dynamically modulate cross-domain interactions, ensuring reliability-modulated feature integration; UGGR constructs a sparse graph for topology-aware relational inference to maintain global semantic and morphological consistency; and UGML dynamically adapts decision margins, preventing the model from overfitting to subjective label noise.
	
	\item \textbf{Uncertainty Alignment (Cols 7-8)}: The superiority of our framework is further evident in the uncertainty maps. UDEL's uncertainty (Col 7) is often overly diffuse, inaccurately thick, or misaligned with the true areas of clinical debate. Unlike UDEL, the uncertainty patterns predicted by UGDD-Net (Col 8) largely correspond to the actual clinical disagreement, effectively capturing diagnostically ambiguous regions. This suggests that our uncertainty estimates hold substantial potential to serve as a trustworthy auxiliary signal to assist dermatologists in human-machine collaborative diagnosis.
\end{itemize}

\section{Discussion}
\label{sec:discussion}

In this study, we presented the Uncertainty-Guided Dual-Domain Network (UGDD-Net), a reliable framework designed to conquer the intrinsic visual ambiguity and morphological irregularity of skin lesions. Moving beyond the limitations of deterministic spatial modeling, our approach fundamentally transforms the role of prediction uncertainty. By explicitly converting uncertainty from a passive, post-hoc evaluation metric into an active guiding signal, UGDD-Net establishes a closed-loop paradigm that enables end-to-end reliability propagation throughout the entire learning lifecycle.

\textbf{Overcoming Deterministic Modeling Bottlenecks:} Powered by a novel two-pass ``glance-and-gaze'' forward mechanism across parallel spatial and spectral pathways, our framework systematically overcomes three critical bottlenecks in existing multi-domain architectures. First, the Uncertainty-Guided Bidirectional Feature Fusion (UGBFF) module employs pixel-level uncertainty to dynamically modulate cross-domain interactions, successfully acting as an intelligent clinical filter to resolve the ``blind fusion'' dilemma. Second, the Uncertainty-Guided Graph Refinement (UGGR) module constructs a priority-guided topological graph to propagate reliable semantic consensus, overcoming ``passive'' reasoning in highly ambiguous regions. Finally, to counteract ``rigid optimization,'' the Uncertainty-Guided Margin-Adaptive Loss (UGML) dynamically adapts decision margins---enforcing strict constraints on confident pixels while relaxing penalties on uncertain ones. This strategy mirrors the cautious deliberation of a human dermatologist, effectively suppressing erratic confidence fluctuations to achieve superior statistical calibration. 

\textbf{Robustness and Clinical Interpretability:} Extensive evaluations across four public benchmarks (ISIC2017, ISIC2018, PH$^2$, and HAM10000) demonstrate that UGDD-Net establishes a new state-of-the-art in segmentation performance. Crucially, its architectural superiority is significantly amplified when evaluated on our curated ``Hard Samples'' subset and under cross-dataset domain shifts, proving its resilience in diverse clinical environments. Beyond exceptional pixel-level accuracy, UGDD-Net delivers profound clinical interpretability. The generated uncertainty maps exhibit diagnostically meaningful alignment with the inter-observer variability of board-certified dermatologists, indicating that the model captures genuine diagnostic ambiguity rather than mere mathematical noise. This clinical relevance highlights the substantial potential of UGDD-Net to evolve from a mere segmentation algorithm into a transparent, trustworthy risk-assessment reference.

\textbf{Limitations and Future Directions:} Despite these promising advancements, several limitations outline critical directions for future research. First, the sophisticated dual-domain architecture and dense topological reasoning inherently increase computational complexity. To facilitate deployment on resource-constrained clinical edge devices, such as mobile dermatoscopes, future efforts will explore knowledge distillation techniques to compress this highly reliable framework into a lightweight student model. Second, our current approach is fundamentally single-modal, relying exclusively on visual pixel data. Because accurate melanoma diagnosis in clinical settings is a holistic process, we plan to evolve UGDD-Net into a multi-modal fusion architecture that integrates dermoscopic images with comprehensive patient metadata, such as genomic history and sequential temporal changes. 

\textbf{Towards Human-Machine Collaborative Diagnosis:} Looking forward, the ultimate goal of AI in dermatology is not to replace the clinician, but to augment their capabilities. We aim to transition UGDD-Net from a standalone, one-shot segmentation algorithm into an interactive, human-in-the-loop clinical system. By utilizing the predicted uncertainty maps as active visual prompts, the system can instinctively guide clinicians to provide targeted feedback (e.g., corrective clicks) specifically in diagnostically ambiguous regions. Dynamically integrating this expert feedback to refine representations in real-time will ultimately evolve our framework into a truly transparent, trustworthy, and collaborative diagnostic partner.

\section{Methods}
\label{sec:methods}

\subsection{Overview}
\label{subsec:overview}

\begin{figure*}[!ht]
	\centering
	\includegraphics[width=1.0\linewidth]{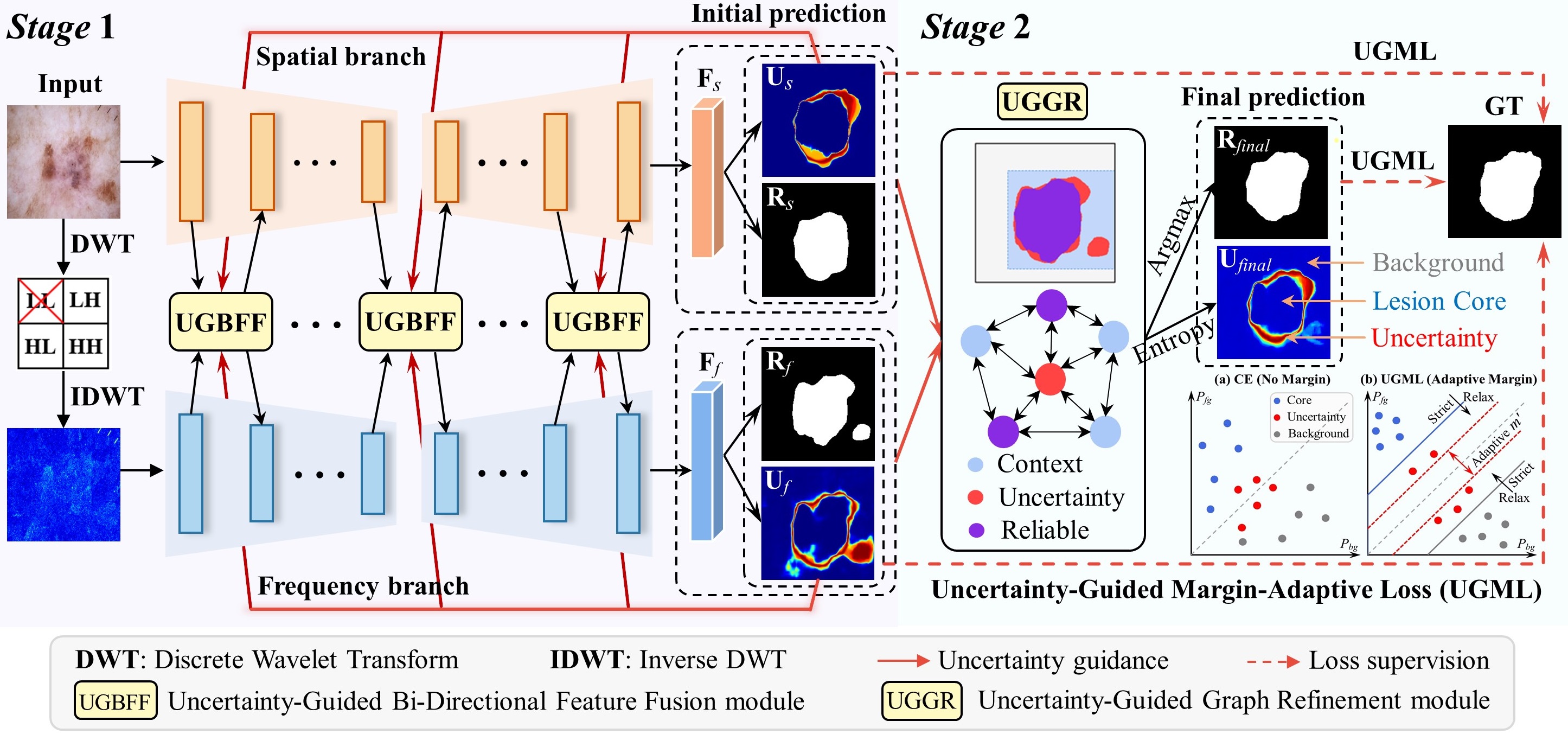}
	\caption{Overview of the proposed Uncertainty-Guided Dual-Domain Network (UGDD-Net). 
		[Structure \& Mechanism]: The architecture employs a parallel dual-stream design: the spatial branch (top, orange) captures semantic context, while the frequency branch (bottom, blue) extracts high-frequency details via Discrete Wavelet Transform (DWT). The entire framework operates under a novel two-pass ``Glance-and-Gaze'' strategy. 
		[Feature Fusion]: Solid red arrows denote the uncertainty guidance flow. Pixel-wise uncertainty maps ($\mathbf{U}_s, \mathbf{U}_f$) generated from Stage 1 are fed back into the Uncertainty-Guided Bi-Directional Feature Fusion (UGBFF) modules. UGBFF implements a reliability-guided rectification mechanism, where high-confidence features, weighted by $1-\mathbf{U}$, from one domain serve as reliable references to rectify the uncertain regions, weighted by $\mathbf{U}$, of the other. 
		[Refinement]: In Stage 2, the Uncertainty-Guided Graph Refinement (UGGR) module optimizes the segmentation by constructing an uncertainty-aware topological graph. Here, reliable consensus and background context nodes actively propagate their semantic priors to guide the representation learning of uncertainty nodes, effectively ensuring global semantic and morphological consistency, yielding the definitive prediction mask ($\mathbf{R}_{final}$) and final uncertainty map ($\mathbf{U}_{final}$). 
		[Supervision]: Dashed red lines indicate supervision by our Uncertainty-Guided Margin-Adaptive Loss (UGML). As illustrated in the bottom right, within the probability space defined by foreground ($P_{fg}$) and background ($P_{bg}$), UGML dynamically applies strict margins to confident pixels to ensure decisive predictions, and relaxed margins to uncertain pixels to prevent overfitting to label noise, ultimately calibrating the model to be highly trustworthy.}
	\label{fig:network_overview}
\end{figure*}

To overcome the aforementioned deterministic bottlenecks---namely, blind fusion, passive reasoning, and rigid optimization---inherent in conventional multi-representation modeling, we propose the Uncertainty-Guided Dual-Domain Network (\textbf{UGDD-Net}, Fig.~\ref{fig:network_overview}). Built upon a parallel spatial and frequency architecture (Sec.~\ref{subsec:backbone}), UGDD-Net transforms predictive uncertainty from a passive posterior metric into an active guiding signal that permeates the entire learning lifecycle. This uncertainty-in-the-loop framework is explicitly driven by three core components: the Uncertainty-Guided Bi-Directional Feature Fusion module (\textbf{UGBFF}, Sec.~\ref{subsec:fusion}), the Uncertainty-Guided Graph Refinement module (\textbf{UGGR}, Sec.~\ref{subsec:graph}), and the Uncertainty-Guided Margin-Adaptive Loss (\textbf{UGML}, Sec.~\ref{subsec:loss}).

To effectively operationalize this closed-loop framework and break the inherent circular dependency of uncertainty estimation (i.e., uncertainty requires predictions, yet refinement requires uncertainty), UGDD-Net is trained via a novel two-pass ``Glance-and-Gaze'' strategy (Sec.~\ref{subsec:training}). This strategy mimics the diagnostic workflow of a dermatologist. The network first performs an initial global ``glance'' to establish a stable semantic representation and quantify uncertainty ($\mathbf{U}_s, \mathbf{U}_f$). Subsequently, an uncertainty-driven ``gaze'' is executed, utilizing these self-generated priors to drive the UGBFF and UGGR modules for active self-rectification, while simultaneously modulating the UGML objective for margin-adaptive supervision.

\subsection{Dual-path Segmentation Framework}
\label{subsec:backbone}

Let $\mathcal{D} = \{(\mathbf{X}_n, \mathbf{Y}_n)\}_{n=1}^N$ denote the dataset, where $\mathbf{X} \in \mathbb{R}^{H \times W \times C}$ represents the input skin lesion image and $\mathbf{Y} \in \{0, 1\}^{H \times W}$ denotes the ground truth binary mask. As illustrated in Fig.~\ref{fig:network_overview}, UGDD-Net processes the input through two parallel pathways---a \textit{spatial branch} and a \textit{frequency branch}---to capture complementary features, followed by a dynamic uncertainty quantification mechanism.

\subsubsection{Spatial Pathway}
This pathway takes the original image as input to extract hierarchical semantic context. To maintain architectural simplicity and establish a robust baseline, we employ the standard U-Net \citep{ronneberger2015u} as the default segmentation backbone $\Phi_s$ (a symmetric architecture is also adopted in the frequency branch). It maps the input image to a high-dimensional deep feature representation $\mathbf{F}_s$, which is subsequently passed through a classification head to produce a dense probability distribution $\mathbf{P}_s$:
\begin{equation}
	\mathbf{F}_s = \Phi_s(\mathbf{X}), \quad \mathbf{P}_s = \sigma(\mathbf{F}_s) \in [0, 1]^{H \times W \times K}
\end{equation}
where $\sigma(\cdot)$ denotes the Softmax activation function, and $K$ is the number of classes (here, $K=2$ for foreground and background). While this pathway excels at capturing overall visual semantics, it often struggles with low-contrast boundaries where spatial gradients are vanishingly weak.

\subsubsection{Frequency Pathway}
To compensate for the spatial limitations, we utilize the frequency domain to explicitly isolate high-frequency boundary details and fine textures. First, we apply the Discrete Wavelet Transform (DWT) to decompose the input image $\mathbf{X}$ into four subbands: one low-frequency approximation ($\mathbf{X}_{LL}$) and three high-frequency details ($\mathbf{X}_{LH}, \mathbf{X}_{HL}, \mathbf{X}_{HH}$). 

Crucially, to force the network to focus strictly on boundary information rather than redundant structural features (which are adequately handled by the Spatial Pathway), we zero out the low-frequency component (i.e., $\mathbf{X}_{LL} = \mathbf{0}$). Subsequently, we apply the Inverse Discrete Wavelet Transform (IDWT) to reconstruct the high-frequency input representation $\mathbf{X}_{freq}$:
\begin{equation}
	\mathbf{X}_{freq} = \text{IDWT}(\mathbf{0}, \mathbf{X}_{LH}, \mathbf{X}_{HL}, \mathbf{X}_{HH})
\end{equation}
Through the IDWT operation, $\mathbf{X}_{freq}$ is restored to the original spatial resolution, i.e., $\mathbf{X}_{freq} \in \mathbb{R}^{H \times W \times C}$. This reconstructed high-frequency image is then processed by the frequency backbone $\Phi_f$ to extract the corresponding frequency-domain feature representation $\mathbf{F}_f$:
\begin{equation}
	\mathbf{F}_f = \Phi_f(\mathbf{X}_{freq}), \quad \mathbf{P}_f = \sigma(\mathbf{F}_f) \in [0, 1]^{H \times W \times K}
\end{equation}

\subsubsection{Prediction and Uncertainty Quantification}
\label{subsubsec:quantification}
For both pathways ($m \in \{s, f\}$), the intermediate feature $\mathbf{F}_m$ yields a dense probability map $\mathbf{P}_m$. To facilitate subsequent graph-based reasoning, we first generate a discrete semantic mask $\mathbf{R}_m$ via the argmax operation:
\begin{equation}
	\mathbf{R}_m^{(i,j)} = \operatorname*{arg\,max}_{k \in \{1, \dots, K\}} \mathbf{P}_m^{(i,j,k)}
\end{equation}

Simultaneously, a reliable measure of confidence is required to drive our closed-loop guidance mechanism. We adopt information entropy to quantify pixel-level prediction uncertainty. To mitigate the ``over-confidence'' trap---where the model may produce low-entropy yet incorrect predictions during the initial \textit{glance} stage---we stabilize the probability estimation using test-time augmentation (TTA). Specifically, for each input image $\mathbf{X}$, we apply a set of $M=4$ spatial geometric transformations $\{\mathcal{T}_v\}_{v=1}^M$ (e.g., rotations and flips) to generate augmented views. After passing each view through the network, we apply the corresponding inverse transformation $\mathcal{T}_v^{-1}$ to spatially realign the predictions before computing the ensemble mean probability $\bar{\mathbf{P}}_m$:
\begin{equation}
	\bar{\mathbf{P}}_m = \frac{1}{M} \sum_{v=1}^{M} \mathcal{T}_v^{-1} \Big( \sigma(\Phi_m(\mathcal{T}_v(\mathbf{X}))) \Big)
\end{equation}
Using this robust distribution, the uncertainty value $\mathbf{U}_m^{(i,j)}$ at pixel $(i,j)$ is computed directly via Information Entropy:
\begin{equation}
	\mathbf{U}_m^{(i,j)} = -\sum_{k=1}^{K} \bar{\mathbf{P}}_m^{(i,j,k)} \log(\bar{\mathbf{P}}_m^{(i,j,k)} + \epsilon)
\end{equation}
where $\epsilon = 10^{-8}$ is a small constant ensuring numerical stability. This TTA-enhanced procedure provides a more stable initial estimation of predictive uncertainty, mitigating the risk of extreme over-confidence caused by deterministic single-pass predictions. The uncertainty map $\mathbf{U}_m$ is then normalized to the range of $[0, 1]$.

\subsection{Uncertainty-Guided Bi-Directional Feature Fusion}
\label{subsec:fusion}

\begin{figure}[!ht]
	\centering
	\includegraphics[width=0.7\linewidth]{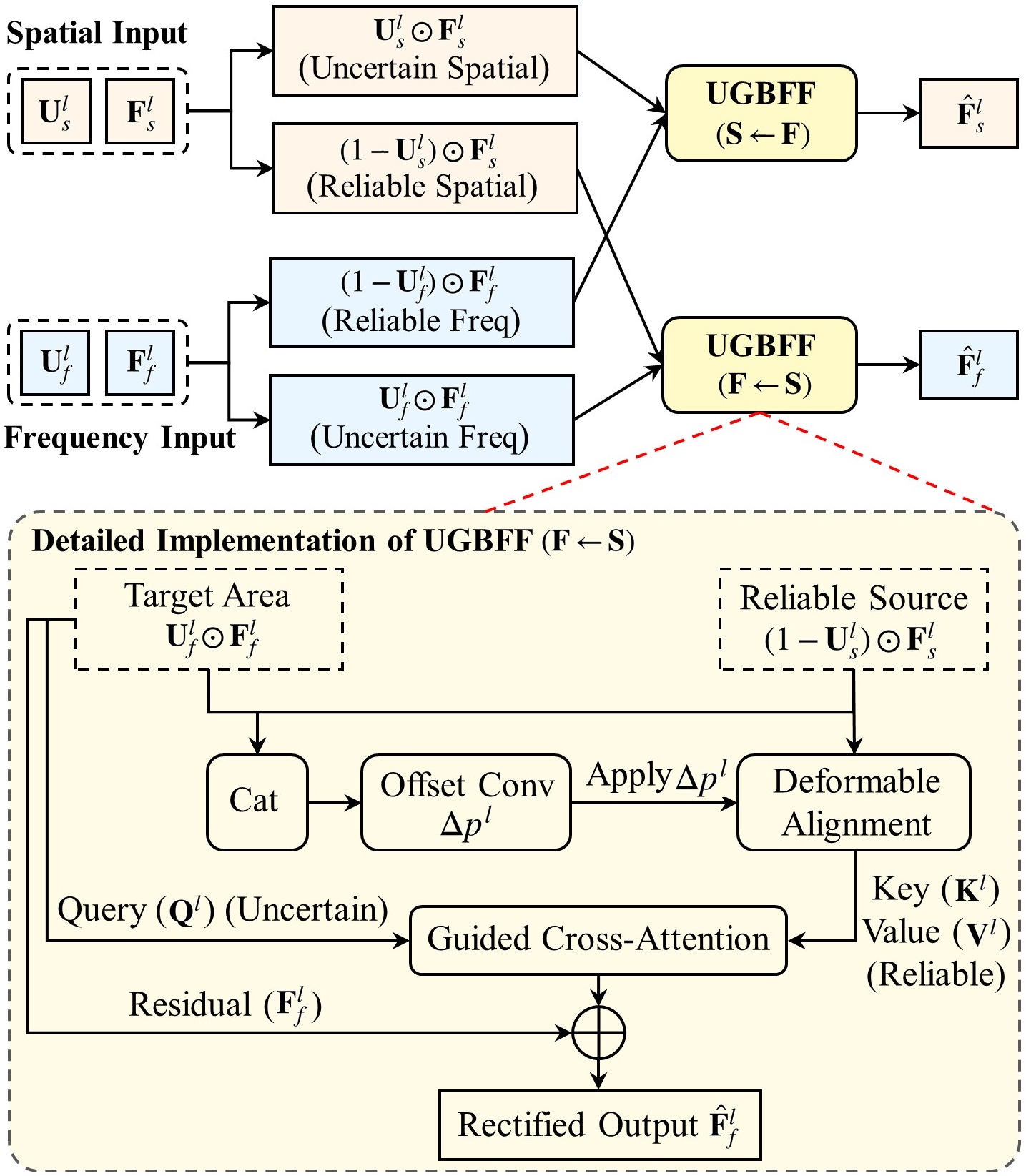}
	\caption{Detailed architecture of the Uncertainty-Guided Bi-Directional Feature Fusion (UGBFF) module. 
		[Bidirectional Formulation]: The top panel illustrates the symmetric cross-domain guidance strategy. Input features at level $l$ ($\smash{\mathbf{F}_s^l}, \smash{\mathbf{F}_f^l}$) are element-wise multiplied by their corresponding rescaled uncertainty maps ($\smash{\mathbf{U}_s^l}, \smash{\mathbf{U}_f^l}$) to isolate uncertain regions (target areas, weighted by $\smash{\mathbf{U}^l}$) and reliable regions (reference sources, weighted by $1-\smash{\mathbf{U}^l}$). 
		[Guided Rectification]: The bottom panel details the implementation of the $\mathbf{F} \leftarrow \mathbf{S}$ fusion process. To address cross-domain spatial misalignment, an offset convolution learns spatial deformations ($\Delta p^l$) to align the reliable spatial source with the target area. Subsequently, a guided cross-attention mechanism dynamically aggregates reliable context, where the uncertain target serves as the Query ($\mathbf{Q}^l$), and the aligned reliable source provides the Key ($\mathbf{K}^l$) and Value ($\mathbf{V}^l$). Finally, a residual connection with the original frequency input ($\smash{\mathbf{F}_f^l}$) produces the rectified output $\smash{\hat{\mathbf{F}}_f^l}$.}
	\label{fig:ugbff}
\end{figure}

To address the limitations of uncertainty-blind interactions prevalent in conventional dual-domain networks, we propose the Uncertainty-Guided Bi-Directional Feature Fusion (UGBFF) module. As illustrated in Fig.~\ref{fig:ugbff}, UGBFF leverages pixel-wise uncertainty maps as active priors to modulate cross-domain information flow. The fusion process is designed to be bidirectional and symmetric; for clarity, we detail the directional rectification from the spatial branch to the frequency branch ($\mathbf{F} \leftarrow \mathbf{S}$). Let $\mathbf{F}_s^l$ and $\mathbf{F}_f^l$ denote the intermediate feature maps at the $l$-th hierarchical level, and $\mathbf{U}_s^l, \mathbf{U}_f^l$ represent the global uncertainty maps ($\mathbf{U}_s, \mathbf{U}_f$) adaptively rescaled to match the spatial resolution of the current level $l$.

\subsubsection{Uncertainty-Aware Feature Selection}
The prerequisite for active rectification is identifying ``what needs correction'' and ``where the reliable knowledge lies.'' Instead of performing global dense attention, we enforce a strict confidence-based gating mechanism. We extract the high-uncertainty regions in the target domain (Frequency) that require refinement, and the low-uncertainty regions in the reference domain (Spatial) to serve as a trustworthy knowledge base. This is formulated via element-wise re-weighting:
\begin{equation}
	\mathbf{F}_{f}^{unc, l} = \mathbf{U}_f^l \odot \mathbf{F}_f^l \quad \text{(Target Area)}
\end{equation}
\begin{equation}
	\mathbf{F}_{s}^{rel, l} = (1 - \mathbf{U}_s^l) \odot \mathbf{F}_s^l \quad \text{(Reliable Source)}
\end{equation}
where $\odot$ denotes element-wise multiplication with channel broadcasting. $\smash{\mathbf{F}_{f}^{unc, l}}$ extracts the ambiguous boundary details requiring rectification, while $\smash{\mathbf{F}_{s}^{rel, l}}$ explicitly suppresses spatial noise to provide a clean, high-confidence semantic reference.


\subsubsection{Deformable Feature Alignment}
Due to the downsampling operations in DWT and inherent domain discrepancies, strict pixel-to-pixel correspondence between the two domains may not hold. To address this spatial misalignment, we employ a deformable alignment strategy prior to cross-attention. 

We first concatenate the uncertain target and the reliable source to learn a spatial offset field $\Delta p^l$:
\begin{equation}
	\Delta p^l = \text{Conv}_{\text{offset}}^l \big( \text{Concat}(\mathbf{F}_{f}^{unc, l}, \mathbf{F}_{s}^{rel, l}) \big)
\end{equation}
These learned offsets $\smash{\Delta p^l}$ are then applied to the reliable spatial source $\smash{\mathbf{F}_{s}^{rel, l}}$ via deformable alignment, resampling the features to spatially align with the frequency domain target:
\begin{equation}
	\mathbf{F}_{s}^{align, l}(p) = \sum_{q} \mathbf{F}_{s}^{rel, l}(q) \cdot \phi(p + \Delta p^l - q)
\end{equation}
where $\phi(\cdot)$ is the bilinear interpolation kernel, $p$ denotes the target grid coordinates, and $q$ enumerates the spatial locations in the source feature.

\subsubsection{Guided Rectification via Cross-Attention}
Finally, we perform feature rectification using a guided cross-attention mechanism. This design enables the uncertain frequency targets to dynamically aggregate reliable context from the aligned spatial sources.

Following the standard cross-attention paradigm, the query ($\smash{\mathbf{Q}^l}$) is projected from the uncertain target $\smash{\mathbf{F}_{f}^{unc, l}}$, while the key ($\smash{\mathbf{K}^l}$) and value ($\smash{\mathbf{V}^l}$) are derived from the aligned reliable source $\smash{\mathbf{F}_{s}^{align, l}}$:
\begin{equation}
	\mathbf{Q}^l = \mathbf{F}_{f}^{unc, l}\mathbf{W}_q^l, \quad \mathbf{K}^l = \mathbf{F}_{s}^{align, l}\mathbf{W}_k^l, \quad \mathbf{V}^l = \mathbf{F}_{s}^{align, l}\mathbf{W}_v^l
\end{equation}
where $\mathbf{W}_q^l, \mathbf{W}_k^l, \mathbf{W}_v^l$ are learnable linear projections for the $l$-th level. By deriving $\smash{\mathbf{K}^l}$ and $\smash{\mathbf{V}^l}$ strictly from uncertainty-filtered features, we guarantee that the frequency branch only retrieves information from trustworthy spatial regions. The attention output is computed as:
\begin{equation}
	\mathbf{F}_{attn}^l = \text{Softmax}\left(\frac{\mathbf{Q}^l(\mathbf{K}^l)^\top}{\sqrt{d_k}} + \mathbf{B}^l\right)\mathbf{V}^l
\end{equation}
where $d_k$ is the scaling factor and $\mathbf{B}^l$ is the relative position bias. To preserve the original frequency characteristics while smoothly integrating the rectified details, the final fused feature $\hat{\mathbf{F}}_f^l$ is obtained via a residual connection with the original frequency input $\mathbf{F}_f^l$:
\begin{equation}
	\hat{\mathbf{F}}_f^l = \mathbf{F}_f^l + \mathbf{F}_{attn}^l
\end{equation}

Symmetrically, the spatial features $\smash{\hat{\mathbf{F}}_s^l}$ are refined by reversing the roles, utilizing reliable frequency details to actively rectify uncertain spatial semantics ($\mathbf{S} \leftarrow \mathbf{F}$).

\textbf{Hierarchical Stage Transition:} It is worth noting that the UGBFF module is designed as a resolution-preserving operation to maintain strict spatial alignment during the cross-attention and residual addition steps. To construct the hierarchical multi-scale architecture (as depicted in Fig.~\ref{fig:network_overview}), the rectified outputs $\smash{\hat{\mathbf{F}}_s^l}$ and $\smash{\hat{\mathbf{F}}_f^l}$ are subsequently processed by stage-specific transition layers. In the encoder, pooling and convolutions are applied to downsample spatial dimensions and adjust channel capacity, whereas transposed convolutions are utilized in the decoder to progressively restore the feature resolution.

\subsection{Uncertainty-Guided Graph Refinement}
\label{subsec:graph}

\begin{figure*}[!ht]
	\centering
	\includegraphics[width=1.0\linewidth]{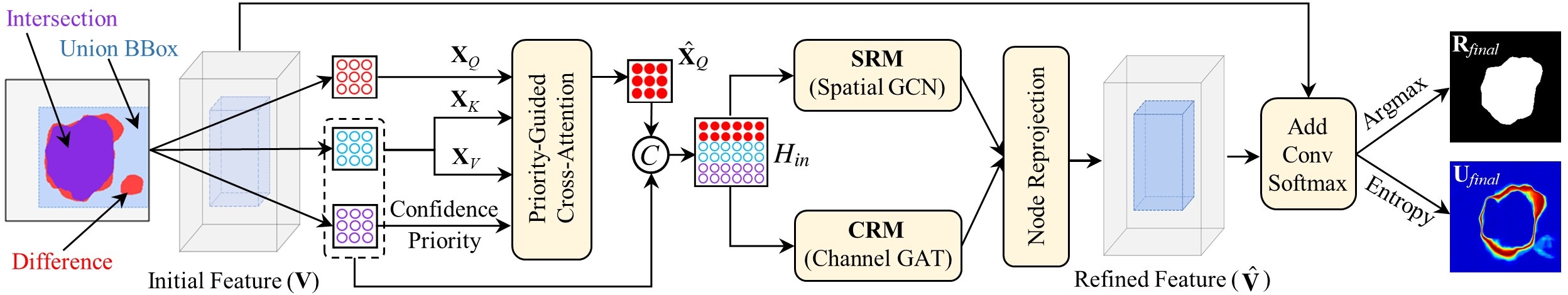}
	\caption{Detailed architecture of the Uncertainty-Guided Graph Refinement (UGGR) module. 
		[Semantic Partitioning]: A composite initial feature tensor is constructed as $\mathbf{V} = \text{Concat}(\mathbf{F}_s, \mathbf{F}_f, \mathbf{U}_s, \mathbf{U}_f)$. Concurrently, a spatial Region of Interest (ROI) is defined by the Union BBox ($\mathbf{R}_s \cup \mathbf{R}_f$, dashed blue line) and partitioned into three mutually exclusive sets: Reliable consensus (Intersection, purple), Uncertainty boundaries (Difference, red), and background Context (remaining area within the Union BBox, blue). 
		[Priority-Guided Interaction]: Graph nodes are explicitly constructed by sampling features from $\mathbf{V}$ according to these spatial partitions. Specifically, Query nodes ($\mathbf{X}_Q$) are sampled exclusively from the Uncertainty region. As denoted by the dashed box, Key/Value reference nodes ($\mathbf{X}_K, \mathbf{X}_V$) are jointly constructed by sampling from both the Context and Reliable regions. A Confidence Priority bias is applied exclusively to the Reliable nodes during cross-attention, actively forcing ambiguous queries to align with the definitive lesion core. 
		[Dual Reasoning \& Reprojection]: The rectified queries ($\hat{\mathbf{X}}_Q$) are concatenated with the reference nodes to form the unified graph ($\mathbf{H}_{in}$), which then undergoes Spatial Reasoning (SRM) and Channel Reasoning (CRM). Finally, a node reprojection operation maps the refined graph back to the continuous spatial grid to generate the refined feature ($\hat{\mathbf{V}}$) and the ultimate predictions ($\mathbf{R}_{final}$, $\mathbf{U}_{final}$).}
	\label{fig:uggr}
\end{figure*}

While the UGBFF module effectively resolves ``blind fusion'' at the local feature level, conventional topological models remain trapped in ``passive reasoning'' by constructing static graphs based merely on raw feature similarity. To dismantle this bottleneck, we propose the Uncertainty-Guided Graph Refinement (UGGR) module (Fig.~\ref{fig:uggr}). Rather than relying on confidence-agnostic dense topologies, UGGR constructs an uncertainty-aware sparse graph. Through a priority-guided cross-attention mechanism, it actively leverages reliable semantic consensus to rectify uncertain query nodes, effectively eliminating foreground-background confusion and ensuring global morphological consistency.

\subsubsection{Semantic Partitioning and Node Construction}
To balance global reasoning capacity with computational tractability, we design a partition-and-sample strategy that focuses exclusively on informative regions. First, we define a localized region of interest (ROI) by extracting the union bounding box of the initial predictions ($\mathbf{R}_s \cup \mathbf{R}_f$). Within this ROI, we construct a multimodal initial feature space $\mathbf{V} = \text{Concat}(\mathbf{F}_s, \mathbf{F}_f, \mathbf{U}_s, \mathbf{U}_f)$, where $\mathbf{F}_s$ and $\mathbf{F}_f$ denote the high-dimensional deep features extracted just before the final classification heads of Stage 1. This ensures that every pixel in $\mathbf{V}$ encapsulates both rich semantic content and explicit reliability priors.

We then partition the ROI into three mutually exclusive semantic sets based on the prediction masks (visualized as colored points in Fig.~\ref{fig:uggr}):
\begin{itemize}
	\item \textbf{Reliable Region} ($\mathcal{R}_{rel}$, purple): The consensus area defined by the intersection $\mathbf{R}_s \cap \mathbf{R}_f$, serving as the trustworthy core.
	\item \textbf{Uncertainty Region} ($\mathcal{R}_{unc}$, red): The ambiguous boundaries defined by the difference $(\mathbf{R}_s \cup \mathbf{R}_f) \setminus \mathcal{R}_{rel}$, requiring active rectification.
	\item \textbf{Context Region} ($\mathcal{R}_{ctx}$, blue): The remaining background defined by $\text{ROI} \setminus (\mathcal{R}_{rel} \cup \mathcal{R}_{unc})$, providing essential global geometric context.
\end{itemize}

From these regions, we sample a fixed number of nodes ($N = 512$) to construct the sparse graph. To avoid redundant node clustering and ensure effective guidance, we employ a hybrid sampling strategy:
\begin{itemize}
	\item \textbf{Query Nodes} ($\mathbf{X}_Q$): We sample $N_{unc} = 256$ pixels with the highest entropy values exclusively from $\mathcal{R}_{unc}$ (red) to form the target nodes requiring rectification.
	\item \textbf{Key/Value Reference Nodes} ($\mathbf{X}_{KV}$): The remaining $N_{ref} = 256$ nodes are jointly sampled from the reliable core ($\mathcal{R}_{rel}$, purple) and the context region ($\mathcal{R}_{ctx}$, blue) using farthest point sampling (FPS). As indicated by the dashed box in Fig.~\ref{fig:uggr}, these two subsets are aggregated to form a unified reference pool for cross-attention. This design mathematically ensures that the query nodes can access both definitive lesion semantics (from purple nodes) and background structural features (from blue nodes).
\end{itemize}

\subsubsection{Priority-Guided Cross-Attention}
To rectify the ambiguous nodes, we model the interaction between $\mathbf{X}_Q$ and the reference nodes $\mathbf{X}_{KV}$ via a priority-guided cross-attention mechanism. Since $\mathbf{X}_{KV}$ contains both lesion core (reliable) and background (context) nodes, we introduce a \textit{confidence priority bias} ($\mathbf{B}_{prior}$) to explicitly guide the attention focus.

For each reference node $j \in \mathbf{X}_{KV}$, we define a static bias scalar $\mathbf{b}_j$:
\begin{equation}
	\mathbf{b}_j = 
	\begin{cases}
		\tau, & \text{if } j \in \mathcal{R}_{rel} \quad (\text{Prioritize Consensus}) \\
		0, & \text{if } j \in \mathcal{R}_{ctx} \quad (\text{Context Reference})
	\end{cases}
\end{equation}
where $\tau$ is a fixed positive constant (empirically set to $1.0$). These scalar biases $\{\mathbf{b}_j\}_{j=1}^{N_{ref}}$ are concatenated to form a bias vector, which is then broadcasted along the query dimension to construct the prior matrix $\mathbf{B}_{prior} \in \mathbb{R}^{N_{unc} \times N_{ref}}$. This matrix directly modulates the attention logits:
\begin{equation}
	\mathbf{A} = \text{Softmax}\left( \frac{(\mathbf{X}_Q \mathbf{W}_Q)(\mathbf{X}_{KV} \mathbf{W}_K)^\top}{\sqrt{d_k}} + \mathbf{B}_{prior} \right)
\end{equation}
The rectified query features are obtained as $\hat{\mathbf{X}}_Q = \mathbf{A} (\mathbf{X}_{KV} \mathbf{W}_V)$. By injecting $\tau$, we structurally amplify the attention weights toward the reliable consensus, forcing the uncertain boundaries to seek semantic alignment with the definitive lesion core, while still keeping the background context accessible.

\subsubsection{Dual Graph Reasoning and Reprojection}
To restore the global topology, we concatenate the rectified uncertainty nodes $\hat{\mathbf{X}}_Q$ with the original reference nodes $\mathbf{X}_{KV}$ along the node dimension to form a unified graph representation $\mathbf{H}_{in} = \text{Concat}_{node}(\hat{\mathbf{X}}_Q, \mathbf{X}_{KV})$. Since these nodes collectively span the entire ROI and retain their original $(x, y)$ spatial coordinates, $\mathbf{H}_{in}$ serves as a complete topological descriptor of the lesion.

Subsequently, we perform spatial-channel graph reasoning:
\begin{itemize}
	\item \textbf{Spatial Reasoning Module (SRM):} Applies a standard graph convolutional network (GCN) \citep{li2022dual} over spatial distances to enforce geometric boundary consistency.
	\item \textbf{Channel Reasoning Module (CRM):} Applies a standard channel-wise graph attention network (GAT) \citep{jiang2024label} to capture global semantic correlations.
\end{itemize}

The outputs from SRM and CRM are aggregated via element-wise summation to produce the final node features. A node reprojection operation is then applied to scatter these discrete graph nodes back to their original pixel coordinates within the continuous ROI, generating a sparse refined tensor $\hat{\mathbf{V}}$. Finally, $\hat{\mathbf{V}}$ is added to the initial ROI features via a residual connection, followed by a convolution layer and a Softmax activation to yield the Stage 2 probability map $\mathbf{P}_{final}$. The definitive segmentation mask $\mathbf{R}_{final}$ and the final uncertainty map $\mathbf{U}_{final}$ are derived from $\mathbf{P}_{final}$ following the identical operations described in Sec.~\ref{subsubsec:quantification}, with the sole distinction that $\mathbf{U}_{final}$ is computed directly via standard information entropy without requiring test-time augmentation.

\subsection{Uncertainty-Guided Margin-Adaptive Loss}
\label{subsec:loss}

\begin{figure}[!ht]
	\centering
	\includegraphics[width=0.7\linewidth]{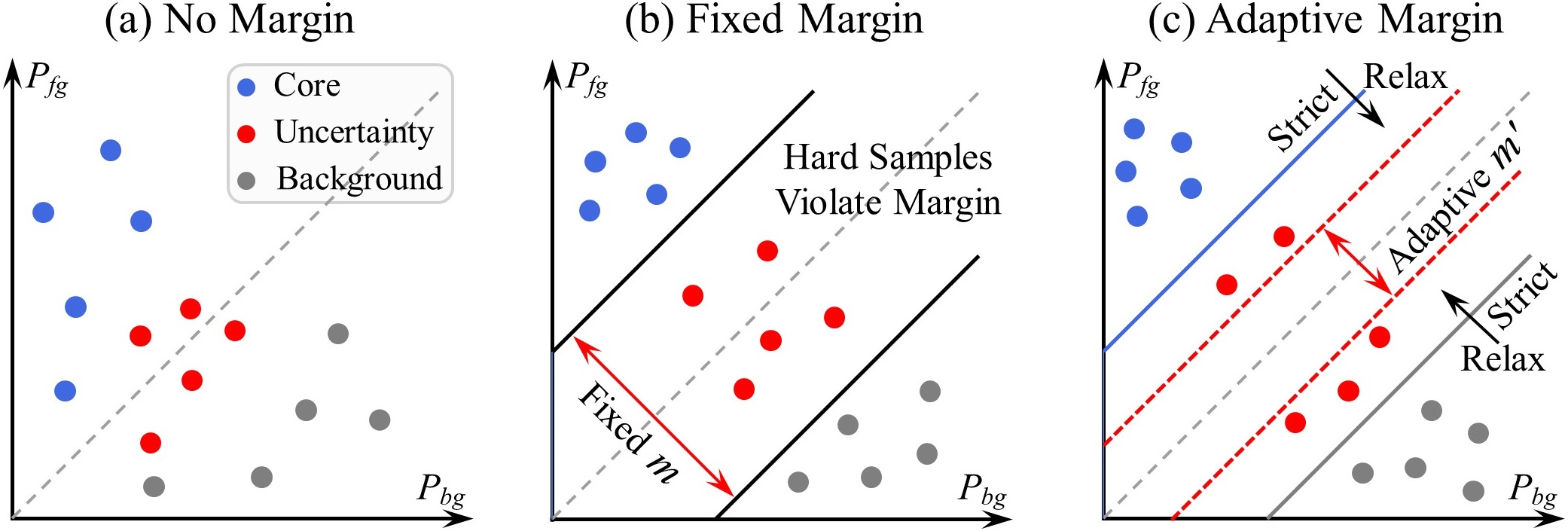} 
	\caption{Conceptual illustration of the Uncertainty-Guided Margin-Adaptive Loss (UGML) in the probability space defined by foreground ($P_{fg}$) and background ($P_{bg}$) predictions. 
		(a) No Margin: Conventional objectives establish a rigid decision boundary ($m=0$), leaving ambiguous features (red points) highly vulnerable to misclassification. 
		(b) Fixed Margin: Enforcing a constant hard margin ($m$) pushes features apart but forces inherently ambiguous samples to violate the margin constraints, causing optimization instability. 
		(c) Adaptive Margin: Our proposed UGML dynamically adjusts the margin boundary based on pixel-wise uncertainty. It enforces strict margins for reliable regions (core, blue; background, grey) to maximize separability, while adaptively relaxing the margin ($m'$) for highly uncertain boundaries (red points) to prevent overfitting to label noise.}
	\label{fig:ugml}
\end{figure}

To break the final bottleneck of ``rigid optimization'' inherent in deterministic networks, we propose the Uncertainty-Guided Margin-Adaptive Loss (UGML). Conventional pixel-wise objectives, such as Cross-Entropy (CE), establish a hard decision boundary without an explicit margin (Fig.~\ref{fig:ugml}(a)). While fixed-margin approaches (Fig.~\ref{fig:ugml}(b)) can enforce feature separability, they impose uncompromising constraints uniformly across all pixels. In skin lesion segmentation, where boundaries are intrinsically ambiguous, aggressively penalizing these uncertain pixels forces the network to overfit subjective label noise, leading to statistical miscalibration and optimization instability.

To resolve this dilemma, UGML acts as an adaptive regularizer that leverages pixel-wise uncertainty to dynamically calibrate the decision boundary. As visualized in Fig.~\ref{fig:ugml}(c), it strictly separates features in \textit{confident regions} while gracefully relaxing the constraints in \textit{ambiguous areas}. For any predicted probability map $\mathbf{P}$ and its corresponding uncertainty map $\mathbf{U}$ generated by our network, the UGML formulation is defined as follows.

\subsubsection{Base Segmentation Objective}
To ensure basic semantic correctness and structural alignment, we employ a widely adopted hybrid segmentation loss $\mathcal{L}_{seg}$. It combines the pixel-wise Cross-Entropy loss ($\mathcal{L}_{CE}$) and the region-based Dice loss ($\mathcal{L}_{Dice}$):
\begin{equation}
	\mathcal{L}_{seg}(\mathbf{P}, \mathbf{Y}) = \mathcal{L}_{CE}(\mathbf{P}, \mathbf{Y}) + \mathcal{L}_{Dice}(\mathbf{P}, \mathbf{Y}),
\end{equation}
where $\mathbf{Y}$ is the ground truth mask.

\subsubsection{Uncertainty-Guided Margin Penalty}
To explicitly handle hard-to-classify pixels, we introduce a dynamic margin ranking formulation. For each pixel $i$, we define an adaptive margin $\xi_i$ modulated by its corresponding uncertainty value $u_i \in \mathbf{U}$:
\begin{equation}
	\xi_i = (1 - u_i) \cdot m,
\end{equation}
where $m$ is a predefined maximum margin scalar (empirically set to $m=0.5$). The margin-adaptive loss is formulated to penalize predictions that fail to maintain this dynamic confidence gap:
\begin{equation}
	\mathcal{L}_{margin}(\mathbf{P}, \mathbf{U}, \mathbf{Y}) = \frac{1}{N} \sum_{i=1}^{N} \max \left( 0, \; \xi_i - (p_{i, gt} - p_{i, non\_gt}) \right).
\end{equation}
Here, $p_{i, gt}$ denotes the predicted probability of the ground-truth class, and $p_{i, non\_gt}$ represents the probability of the non-target class. In our binary lesion segmentation task, these dynamically correspond to the foreground probability ($P_{fg}$) and background probability ($P_{bg}$) depending on the pixel's true label. The term $(p_{i, gt} - p_{i, non\_gt})$ intrinsically quantifies the \textit{confidence gap}.

\textbf{Mechanism Analysis:} The adaptive term $\xi_i$ enables the loss function to dynamically transition between two distinct optimization modes, mirroring the visual intuition in Fig.~\ref{fig:ugml}(c):
\begin{itemize}
	\item \textbf{Strict Constraint} (blue/grey points, $u_i \to 0$): In reliable regions, the effective margin $\xi_i$ approaches the maximum value $m$. This heavily penalizes the network unless it produces a decisively confident prediction (i.e., widening the gap between $P_{fg}$ and $P_{bg}$), explicitly maximizing feature separability for the definitive lesion core and clear background.
	\item \textbf{Relaxed Constraint} (red points, $u_i \to 1$): In highly ambiguous boundary regions, predictions often cluster near the decision boundary ($P_{fg} \approx P_{bg}$). Here, $\xi_i$ adaptively shrinks toward zero. This mathematical tolerance prevents the model from aggressively overfitting to subjective label noise, ensuring it remains appropriately cautious and effectively circumventing the ``margin violation'' collapse.
\end{itemize}

\subsubsection{UGML Formulation}
The complete UGML objective for a given prediction is the weighted integration of the base segmentation loss and the adaptive margin penalty:
\begin{equation}
	\mathcal{L}_{UGML}(\mathbf{P}, \mathbf{U}, \mathbf{Y}) = \mathcal{L}_{seg}(\mathbf{P}, \mathbf{Y}) + \lambda \cdot \mathcal{L}_{margin}(\mathbf{P}, \mathbf{U}, \mathbf{Y}).
\end{equation}
Empirically, we set the weighting coefficient $\lambda = 0.1$ to ensure stable convergence without overshadowing the primary semantic segmentation objective.

\subsection{Network Training and Inference Strategy}
\label{subsec:training}

The proposed UGDD-Net introduces an \textit{inherent circular dependency} during the forward pass: the uncertainty-guided bi-directional feature fusion (UGBFF) modules require the global uncertainty maps ($\mathbf{U}_s, \mathbf{U}_f$) as driving signals, yet these maps are only computed from the final decoder outputs. To break this causal loop, we design a two-pass ``glance-and-gaze'' forward mechanism coupled with a two-stage curriculum training strategy.

\textbf{Two-Pass Forward Mechanism:} To obtain the requisite uncertainty priors before performing active refinement, each input undergoes two sequential forward passes during a single iteration. In \textbf{Pass 1 (Glance)}, the input is initially processed with the active guidance in all UGBFF modules temporarily disabled ($\mathbf{U} \equiv \mathbf{0}$). This rapid ``glance'' establishes a foundational cross-domain representation, generating the initial probability maps ($\mathbf{P}_s^{(1)}, \mathbf{P}_f^{(1)}$) and their corresponding uncertainty distributions ($\mathbf{U}_s, \mathbf{U}_f$) via information entropy. In \textbf{Pass 2 (Gaze)}, the generated $\mathbf{U}_s$ and $\mathbf{U}_f$ are fed back into all UGBFF modules to actively gate the feature flow. The rectified features then undergo UGGR topological reasoning, yielding the definitive prediction $\mathbf{P}_{final}$ and $\mathbf{U}_{final}$.

\textbf{Two-Stage Curriculum Training:} In the early epochs, utilizing uncalibrated uncertainty noise from immature predictions to drive the active modules would inevitably corrupt feature learning. Therefore, we implement a curriculum paradigm:

\textbf{Stage 1 (Base Initialization):} The backbone is optimized purely on the ``glance'' pass outputs using the base segmentation objective against the ground truth $\mathbf{Y}$:
\begin{equation}
	\mathcal{L}_{stage1} = \mathcal{L}_{seg}(\mathbf{P}_s^{(1)}, \mathbf{Y}) + \mathcal{L}_{seg}(\mathbf{P}_f^{(1)}, \mathbf{Y}).
\end{equation}
Once the validation loss plateaus (controlled by a patience hyperparameter of 20 epochs), the base semantic representation is deemed stable, and training transitions to Stage 2.

\textbf{Stage 2 (Uncertainty-Reliability Calibration):} The complete ``glance-and-gaze'' mechanism is activated, leveraging the stabilized uncertainty maps for active routing. To comprehensively calibrate the network, we apply the uncertainty-guided margin-adaptive loss (UGML) in a deep supervision manner, jointly optimizing the intermediate dual-domain outputs and the final prediction:
\begin{equation}
	\begin{split}
		\mathcal{L}_{stage2} &= \mathcal{L}_{UGML}(\mathbf{P}_s^{(1)}, \mathbf{U}_s, \mathbf{Y}) \\
		&\quad + \mathcal{L}_{UGML}(\mathbf{P}_f^{(1)}, \mathbf{U}_f, \mathbf{Y}) \\
		&\quad + \mathcal{L}_{UGML}(\mathbf{P}_{final}, \mathbf{U}_{final}, \mathbf{Y}).
	\end{split}
\end{equation}

\textbf{Inference Phase:} During testing, network weights are frozen and the curriculum optimization is omitted. For each unseen image, the model directly executes the complete two-pass ``glance-and-gaze'' mechanism, deriving the ultimate segmentation mask strictly from the refined $\mathbf{P}_{final}$ generated in the second pass.

\vspace{1em}
\backmatter

\bmhead{Data availability}
The datasets analyzed during the current study are publicly available. The ISIC2017 and ISIC2018 datasets can be accessed at the official ISIC archive (\url{https://challenge.isic-archive.com}). The PH$^2$ dataset is available at \url{https://www.fc.up.pt/addi/ph2\%20database.html}. The HAM10000 dataset can be accessed via the Harvard Dataverse (\url{https://doi.org/10.7910/DVN/DBW86T}).

\bmhead{Code availability}
The source code and associated scripts supporting the findings of this study will be made publicly available upon publication of the manuscript. During the peer-review process, the code is available to the editors and reviewers upon reasonable request.

\bmhead{Author contributions}
D.D. and W.Z. conceived the study and designed the network architecture. D.D., C.D., G.D., and Q.Y. implemented the code and performed the experiments. Q.Z., P.R., and G.K. provided clinical insights and performed the clinical interpretability analysis. G.K. and W.Z. supervised the project and provided critical revisions to the manuscript. All authors reviewed and approved the final manuscript.

\bmhead{Competing interests}
The authors declare no competing interests.

\bmhead{Acknowledgments}
This work was supported by the National Natural Science Foundation of China (Grant Nos. 82302309, 62301413, 12090021, and 81971766), the Sichuan Provincial Natural Science Foundation (Grant No. 2025ZNSFSC0640), and the China Postdoctoral Science Foundation (Grant No. 2021M692577).

	

\begin{thebibliography}{10}
\expandafter\ifx\csname url\endcsname\relax
  \def\url#1{\burl{#1}}\fi
\expandafter\ifx\csname urlprefix\endcsname\relax\def\urlprefix{URL }\fi
\providecommand{\bibinfo}[2]{#2}
\providecommand{\eprint}[2][]{\url{#2}}
\providecommand{\doi}[1]{\url{https://doi.org/#1}}
\bibcommenthead

\bibitem{langselius2025global}
\bibinfo{author}{Langselius, O.} \emph{et~al.}
\newblock \bibinfo{title}{Global burden of cutaneous melanoma incidence
  attributable to ultraviolet radiation in 2022}.
\newblock \emph{\bibinfo{journal}{International Journal of Cancer}}
  \textbf{\bibinfo{volume}{157}}, \bibinfo{pages}{1110--1119}
  (\bibinfo{year}{2025}).

\bibitem{hrvatin2026completed}
\bibinfo{author}{Hrvatin~Stancic, B.} \emph{et~al.}
\newblock \bibinfo{title}{Completed suicide in patients with skin disease: a
  systematic review and meta-analysis}.
\newblock \emph{\bibinfo{journal}{Journal of the European Academy of
  Dermatology and Venereology}} \textbf{\bibinfo{volume}{40}},
  \bibinfo{pages}{46--58} (\bibinfo{year}{2026}).

\bibitem{celebi2019dermoscopy}
\bibinfo{author}{Celebi, M.~E.}, \bibinfo{author}{Codella, N.} \&
  \bibinfo{author}{Halpern, A.}
\newblock \bibinfo{title}{Dermoscopy image analysis: overview and future
  directions}.
\newblock \emph{\bibinfo{journal}{IEEE journal of biomedical and health
  informatics}} \textbf{\bibinfo{volume}{23}}, \bibinfo{pages}{474--478}
  (\bibinfo{year}{2019}).

\bibitem{wang2025udel}
\bibinfo{author}{Wang, K.}, \bibinfo{author}{Liu, J.}, \bibinfo{author}{An, Z.}
  \& \bibinfo{author}{Lu, Y.}
\newblock \bibinfo{title}{Udel: Rethinking uncertainty dynamic estimation
  learning for ambiguous medical image segmentation}.
\newblock \emph{\bibinfo{journal}{Digital Signal Processing}}
  \bibinfo{pages}{105723} (\bibinfo{year}{2025}).

\bibitem{ronneberger2015u}
\bibinfo{author}{Ronneberger, O.}, \bibinfo{author}{Fischer, P.} \&
  \bibinfo{author}{Brox, T.}
\newblock \emph{\bibinfo{title}{U-net: Convolutional networks for biomedical
  image segmentation}}, \bibinfo{pages}{234--241}
  (\bibinfo{publisher}{Springer}, \bibinfo{year}{2015}).

\bibitem{zhou2022h}
\bibinfo{author}{Zhou, X.} \emph{et~al.}
\newblock \bibinfo{title}{H-net: A dual-decoder enhanced fcnn for automated
  biomedical image diagnosis}.
\newblock \emph{\bibinfo{journal}{Information Sciences}}
  \textbf{\bibinfo{volume}{613}}, \bibinfo{pages}{575--590}
  (\bibinfo{year}{2022}).

\bibitem{dai2025improving}
\bibinfo{author}{Dai, D.} \emph{et~al.}
\newblock \bibinfo{title}{Improving the performance of medical image
  segmentation with instructive feature learning}.
\newblock \emph{\bibinfo{journal}{Medical Image Analysis}}
  \bibinfo{pages}{103818} (\bibinfo{year}{2025}).

\bibitem{cao2022swin}
\bibinfo{author}{Cao, H.} \emph{et~al.}
\newblock \emph{\bibinfo{title}{Swin-unet: Unet-like pure transformer for
  medical image segmentation}}, \bibinfo{pages}{205--218}
  (\bibinfo{publisher}{Springer}, \bibinfo{year}{2022}).

\bibitem{ji2025bdformer}
\bibinfo{author}{Ji, Z.}, \bibinfo{author}{Ye, Y.} \& \bibinfo{author}{Ma, X.}
\newblock \bibinfo{title}{Bdformer: Boundary-aware dual-decoder transformer for
  skin lesion segmentation}.
\newblock \emph{\bibinfo{journal}{Artificial Intelligence in Medicine}}
  \textbf{\bibinfo{volume}{162}}, \bibinfo{pages}{103079}
  (\bibinfo{year}{2025}).

\bibitem{ruan2024vm}
\bibinfo{author}{Ruan, J.}, \bibinfo{author}{Li, J.} \& \bibinfo{author}{Xiang,
  S.}
\newblock \bibinfo{title}{Vm-unet: Vision mamba unet for medical image
  segmentation}.
\newblock \emph{\bibinfo{journal}{ACM Transactions on Multimedia Computing,
  Communications and Applications}}  (\bibinfo{year}{2024}).

\bibitem{liu2025vision}
\bibinfo{author}{Liu, X.} \emph{et~al.}
\newblock \bibinfo{title}{Vision mamba: A comprehensive survey and taxonomy}.
\newblock \emph{\bibinfo{journal}{IEEE Transactions on Neural Networks and
  Learning Systems}}  (\bibinfo{year}{2025}).

\bibitem{liu2025sfma}
\bibinfo{author}{Liu, Z.}, \bibinfo{author}{Zhang, Y.}, \bibinfo{author}{Wang,
  B.}, \bibinfo{author}{Yang, Y.} \& \bibinfo{author}{Cai, L.}
\newblock \emph{\bibinfo{title}{Sfma-unet: A mamba-based spatial-frequency
  fusion network for medical image segmentation}}, \bibinfo{pages}{1--5}
  (\bibinfo{publisher}{IEEE}, \bibinfo{year}{2025}).

\bibitem{cai2026cdmt}
\bibinfo{author}{Cai, Y.}, \bibinfo{author}{Liu, Y.}, \bibinfo{author}{Li, J.}
  \& \bibinfo{author}{Xiong, Q.}
\newblock \bibinfo{title}{Cdmt-unet: a dual-frequency cross-fusion model for
  skin cancer segmentation}.
\newblock \emph{\bibinfo{journal}{Biomedical Signal Processing and Control}}
  \textbf{\bibinfo{volume}{113}}, \bibinfo{pages}{108880}
  (\bibinfo{year}{2026}).

\bibitem{rahman2024g}
\bibinfo{author}{Rahman, M.~M.} \& \bibinfo{author}{Marculescu, R.}
\newblock \emph{\bibinfo{title}{{G-Cascade}: Efficient cascaded graph
  convolutional decoding for {2D} medical image segmentation}},
  \bibinfo{pages}{7728--7737} (\bibinfo{year}{2024}).

\bibitem{kui2025wingraphunet}
\bibinfo{author}{Kui, X.} \emph{et~al.}
\newblock \bibinfo{title}{Wingraphunet: Advanced windowed graph modeling with
  remixed contextual learning for efficient medical image segmentation}.
\newblock \emph{\bibinfo{journal}{Knowledge-Based Systems}}
  \bibinfo{pages}{114417} (\bibinfo{year}{2025}).

\bibitem{zou2025toward}
\bibinfo{author}{Zou, K.} \emph{et~al.}
\newblock \bibinfo{title}{Toward reliable medical image segmentation by
  modeling evidential calibrated uncertainty}.
\newblock \emph{\bibinfo{journal}{IEEE Transactions on Cybernetics}}
  (\bibinfo{year}{2025}).

\bibitem{munia2025attention}
\bibinfo{author}{Munia, A.~A.} \emph{et~al.}
\newblock \bibinfo{title}{Attention-guided hierarchical fusion u-net for
  uncertainty-driven medical image segmentation}.
\newblock \emph{\bibinfo{journal}{Information Fusion}}
  \textbf{\bibinfo{volume}{115}}, \bibinfo{pages}{102719}
  (\bibinfo{year}{2025}).

\bibitem{zhou2024spatial}
\bibinfo{author}{Zhou, Z.} \emph{et~al.}
\newblock \emph{\bibinfo{title}{Spatial-frequency dual domain attention network
  for medical image segmentation}}, \bibinfo{pages}{4076--4081}
  (\bibinfo{publisher}{IEEE}, \bibinfo{year}{2024}).

\bibitem{huang2025dpmf}
\bibinfo{author}{Huang, Y.} \emph{et~al.}
\newblock \bibinfo{title}{Dpmf-net: A dual-path perceptive multi-stage fusion
  network for skin lesion segmentation}.
\newblock \emph{\bibinfo{journal}{Engineering Applications of Artificial
  Intelligence}} \textbf{\bibinfo{volume}{161}}, \bibinfo{pages}{112043}
  (\bibinfo{year}{2025}).

\bibitem{li2026learning}
\bibinfo{author}{Li, X.} \emph{et~al.}
\newblock \bibinfo{title}{Learning geometric and visual features for medical
  image segmentation with vision gnn}.
\newblock \emph{\bibinfo{journal}{Computerized Medical Imaging and Graphics}}
  \bibinfo{pages}{102720} (\bibinfo{year}{2026}).

\bibitem{shore2003properties}
\bibinfo{author}{Shore, J.} \& \bibinfo{author}{Johnson, R.}
\newblock \bibinfo{title}{Properties of cross-entropy minimization}.
\newblock \emph{\bibinfo{journal}{IEEE Transactions on Information Theory}}
  \textbf{\bibinfo{volume}{27}}, \bibinfo{pages}{472--482}
  (\bibinfo{year}{2003}).

\bibitem{milletari2016v}
\bibinfo{author}{Milletari, F.}, \bibinfo{author}{Navab, N.} \&
  \bibinfo{author}{Ahmadi, S.-A.}
\newblock \emph{\bibinfo{title}{V-net: Fully convolutional neural networks for
  volumetric medical image segmentation}}, \bibinfo{pages}{565--571}
  (\bibinfo{publisher}{Ieee}, \bibinfo{year}{2016}).

\bibitem{lin2017focal}
\bibinfo{author}{Lin, T.-Y.}, \bibinfo{author}{Goyal, P.},
  \bibinfo{author}{Girshick, R.}, \bibinfo{author}{He, K.} \&
  \bibinfo{author}{Doll{\'a}r, P.}
\newblock \emph{\bibinfo{title}{{Focal Loss} for dense object detection}},
  \bibinfo{pages}{2980--2988} (\bibinfo{year}{2017}).

\bibitem{kervadec2019boundary}
\bibinfo{author}{Kervadec, H.} \emph{et~al.}
\newblock \emph{\bibinfo{title}{Boundary loss for highly unbalanced
  segmentation}}, \bibinfo{pages}{285--296} (\bibinfo{publisher}{PMLR},
  \bibinfo{year}{2019}).

\bibitem{codella2018skin}
\bibinfo{author}{Codella, N.~C.} \emph{et~al.}
\newblock \emph{\bibinfo{title}{Skin lesion analysis toward melanoma detection:
  A challenge at the 2017 international symposium on biomedical imaging (isbi),
  hosted by the international skin imaging collaboration (isic)}},
  \bibinfo{pages}{168--172} (\bibinfo{publisher}{IEEE}, \bibinfo{year}{2018}).

\bibitem{bissoto2018deep}
\bibinfo{author}{Bissoto, A.} \emph{et~al.}
\newblock \bibinfo{title}{Deep-learning ensembles for skin-lesion segmentation,
  analysis, classification: Recod titans at isic challenge 2018}.
\newblock \emph{\bibinfo{journal}{arXiv preprint arXiv:1808.08480}}
  (\bibinfo{year}{2018}).

\bibitem{mendoncca2013ph}
\bibinfo{author}{Mendon{\c{c}}a, T.}, \bibinfo{author}{Ferreira, P.~M.},
  \bibinfo{author}{Marques, J.~S.}, \bibinfo{author}{Marcal, A.~R.} \&
  \bibinfo{author}{Rozeira, J.}
\newblock \emph{\bibinfo{title}{Ph 2-a dermoscopic image database for research
  and benchmarking}}, \bibinfo{pages}{5437--5440} (\bibinfo{publisher}{IEEE},
  \bibinfo{year}{2013}).

\bibitem{tschandl2018ham10000}
\bibinfo{author}{Tschandl, P.}, \bibinfo{author}{Rosendahl, C.} \&
  \bibinfo{author}{Kittler, H.}
\newblock \bibinfo{title}{The ham10000 dataset, a large collection of
  multi-source dermatoscopic images of common pigmented skin lesions}.
\newblock \emph{\bibinfo{journal}{Scientific data}}
  \textbf{\bibinfo{volume}{5}}, \bibinfo{pages}{1--9} (\bibinfo{year}{2018}).

\bibitem{altman2017points}
\bibinfo{author}{Altman, N.} \& \bibinfo{author}{Krzywinski, M.}
\newblock \bibinfo{title}{Points of significance: interpreting p values}.
\newblock \emph{\bibinfo{journal}{Nature methods}}
  \textbf{\bibinfo{volume}{14}}, \bibinfo{pages}{213--215}
  (\bibinfo{year}{2017}).

\bibitem{huang2025lesion}
\bibinfo{author}{Huang, X.} \emph{et~al.}
\newblock \bibinfo{title}{Lesion boundary detection for skin lesion
  segmentation based on boundary sensing and cnn-transformer fusion networks}.
\newblock \emph{\bibinfo{journal}{Artificial Intelligence in Medicine}}
  \bibinfo{pages}{103190} (\bibinfo{year}{2025}).

\bibitem{zou2024skinmamba}
\bibinfo{author}{Zou, S.}, \bibinfo{author}{Zhang, M.}, \bibinfo{author}{Fan,
  B.}, \bibinfo{author}{Zhou, Z.} \& \bibinfo{author}{Zou, X.}
\newblock \bibinfo{title}{Skinmamba: A precision skin lesion segmentation
  architecture with cross-scale global state modeling and frequency boundary
  guidance}.
\newblock \emph{\bibinfo{journal}{arXiv preprint arXiv:2409.10890}}
  (\bibinfo{year}{2024}).

\bibitem{wang2025dpgnet}
\bibinfo{author}{Wang, H.} \emph{et~al.}
\newblock \bibinfo{title}{Dpgnet: A boundary-aware medical image segmentation
  framework via uncertainty perception}.
\newblock \emph{\bibinfo{journal}{IEEE Journal of Biomedical and Health
  Informatics}}  (\bibinfo{year}{2025}).

\bibitem{li2026cfformer}
\bibinfo{author}{Li, J.} \emph{et~al.}
\newblock \bibinfo{title}{Cfformer: Cross cnn-transformer channel attention and
  spatial feature fusion for improved segmentation of heterogeneous medical
  images}.
\newblock \emph{\bibinfo{journal}{Expert Systems with Applications}}
  \textbf{\bibinfo{volume}{295}}, \bibinfo{pages}{128835}
  (\bibinfo{year}{2026}).

\bibitem{zhong2025dsu}
\bibinfo{author}{Zhong, L.} \emph{et~al.}
\newblock \bibinfo{title}{Dsu-net: Dual-stage u-net based on cnn and
  transformer for skin lesion segmentation}.
\newblock \emph{\bibinfo{journal}{Biomedical Signal Processing and Control}}
  \textbf{\bibinfo{volume}{100}}, \bibinfo{pages}{107090}
  (\bibinfo{year}{2025}).

\bibitem{zhai2024dma}
\bibinfo{author}{Zhai, G.}, \bibinfo{author}{Wang, G.}, \bibinfo{author}{Shang,
  Q.}, \bibinfo{author}{Li, Y.} \& \bibinfo{author}{Wang, H.}
\newblock \bibinfo{title}{Dma-net: A dual branch encoder and multi-scale cross
  attention fusion network for skin lesion segmentation}.
\newblock \emph{\bibinfo{journal}{IET Image Processing}}
  \textbf{\bibinfo{volume}{18}}, \bibinfo{pages}{4531--4541}
  (\bibinfo{year}{2024}).

\bibitem{zhou2025f2cau}
\bibinfo{author}{Zhou, T.} \emph{et~al.}
\newblock \bibinfo{title}{F2cau-net: A dual fuzzy medical image segmentation
  cascade method based on fuzzy feature learning}.
\newblock \emph{\bibinfo{journal}{Applied Soft Computing}}
  \bibinfo{pages}{113692} (\bibinfo{year}{2025}).

\bibitem{li2022dual}
\bibinfo{author}{Li, Y.} \emph{et~al.}
\newblock \bibinfo{title}{Dual encoder-based dynamic-channel graph
  convolutional network with edge enhancement for retinal vessel segmentation}.
\newblock \emph{\bibinfo{journal}{IEEE Transactions on Medical Imaging}}
  \textbf{\bibinfo{volume}{41}}, \bibinfo{pages}{1975--1989}
  (\bibinfo{year}{2022}).

\bibitem{jiang2024label}
\bibinfo{author}{Jiang, Q.}, \bibinfo{author}{Ye, H.}, \bibinfo{author}{Yang,
  B.} \& \bibinfo{author}{Cao, F.}
\newblock \bibinfo{title}{Label-decoupled medical image segmentation with
  spatial-channel graph convolution and dual attention enhancement}.
\newblock \emph{\bibinfo{journal}{IEEE journal of biomedical and health
  informatics}} \textbf{\bibinfo{volume}{28}}, \bibinfo{pages}{2830--2841}
  (\bibinfo{year}{2024}).

\end{thebibliography}

\end{document}